%% file: main.tex
\documentclass[fleqn,usenatbib]{mnras}

\usepackage{newtxtext,newtxmath}

\usepackage[T1]{fontenc}

\DeclareRobustCommand{\VAN}[3]{#2}
\let\VANthebibliography\thebibliography
\def\thebibliography{\DeclareRobustCommand{\VAN}[3]{##3}\VANthebibliography}

\usepackage{graphicx}	
\usepackage{amsmath}	
\usepackage{subcaption}
\usepackage{lineno}
\usepackage{graphicx}

\input{commands}

\input{hyperlink-year-only-natbib-patch}

\graphicspath{{./figures/}{./}}

\title[Disrupting satellites in Auriga]{Auriga Streams I: disrupting satellites surrounding Milky Way-mass haloes at multiple resolutions}

\author[A.H.~Riley et al.]{Alexander H.~Riley,$^{1}$\thanks{E-mail: alexander.riley2@durham.ac.uk}
Nora Shipp,$^{2}$
Christine M.~Simpson,$^{3}$
Rebekka Bieri,$^{4}$
Azadeh Fattahi,$^{1}$ \newauthor
Shaun T.~Brown,$^{1}$
Kyle A. Oman,$^{1,5}$
Francesca Fragkoudi,$^{1}$
Facundo A.~G\'{o}mez,$^{6}$
Robert J.~J.~Grand,$^{7}$ \newauthor
and Federico Marinacci$^{8,9}$
\\
$^{1}$Institute for Computational Cosmology, Department of Physics, Durham University, South Road, Durham DH1 3LE, UK\\
$^{2}$Department of Astronomy, University of Washington, Seattle, WA 98195, USA\\
$^{3}$Argonne Leadership Computing Facility, Argonne National Laboratory, Lemont, IL 60439, USA\\
$^{4}$Department of Astrophysics, University of Zurich, 8057 Zurich, Switzerland\\
$^{5}$Centre for Extragalactic Astronomy, Department of Physics, Durham University, South Road, Durham DH1 3LE, UK\\
$^{6}$Departamento de Astronom\'{i}a, Universidad de La Serena, Av.~Juan Cisternas 1200 Norte, La Serena, Chile\\
$^{7}$Astrophysics Research Institute, Liverpool John Moores University, 146 Brownlow Hill, Liverpool, L3 5RF, UK\\
$^{8}$Department of Physics \& Astronomy `Augusto Righi', University of Bologna, via Gobetti 93/2, I-40129 Bologna, Italy\\
$^{9}$INAF, Astrophysics and Space Science Observatory Bologna, Via P. Gobetti 93/3, I-40129 Bologna, Italy
}

\date{Accepted 2025 August 12. Received 2025 July 8; in original form 2024 October 21}

\pubyear{2025}

\begin{document}
\label{firstpage}
\pagerange{\pageref{firstpage}--\pageref{lastpage}}
\maketitle

\begin{abstract}
    In a hierarchically formed Universe, galaxies accrete smaller systems that tidally disrupt as they evolve in the host's potential.
    We present a complete catalogue of disrupting galaxies accreted onto Milky Way-mass haloes from the Auriga suite of cosmological magnetohydrodynamic zoom-in simulations.
    We classify accretion events as intact satellites, stellar streams, or phase-mixed systems based on automated criteria calibrated to a visually classified sample, and match accretions to their counterparts in haloes re-simulated at higher resolution.
    Most satellites at the present day have lost substantial amounts of stellar mass -- 67~per~cent have $f_\text{bound} < 0.97$ (our threshold of lost stellar mass to no longer be considered intact), while 53~per~cent satisfy a more stringent $f_\text{bound} < 0.8$.
    Streams typically outnumber intact systems, contribute a smaller fraction of overall accreted stars, and are substantial contributors at intermediate distances from the host centre ($\sim$0.1 to $\sim$0.7$R_\text{200m}$, or $\sim$35 to $\sim$250~kpc for the Milky Way).
    We also identify accretion events that disrupt to form streams around massive intact satellites instead of the main host.
    Streams are more likely than intact or phase-mixed systems to have experienced preprocessing, suggesting this mechanism is important for setting disruption rates around Milky Way-mass haloes.
    All of these results are preserved across different simulation resolutions, though we do find some hints that satellites disrupt more readily at lower resolution.
    The Auriga haloes suggest that disrupting satellites surrounding Milky Way-mass galaxies are the norm and that a wealth of tidal features waits to be uncovered in upcoming surveys.
\end{abstract}




\section{Introduction}

In both the real Universe and the prevailing cosmological paradigm -- cold dark matter with a cosmological constant ($\Lambda$CDM) -- structure forms hierarchically \citep{Cole:2000} resulting in galaxies surrounded by stellar haloes formed by the remnants of smaller accretions \citep{Searle:1978}.
As smaller galaxies fall into and orbit around a more massive host, they begin to unravel and form extended stellar streams that eventually phase-mix into a spatially smooth halo \citep{Bullock:2005, Cooper:2010, Gomez:2013}.

\begin{figure*}
    \begin{center}
    \includegraphics[width=1.0\linewidth]{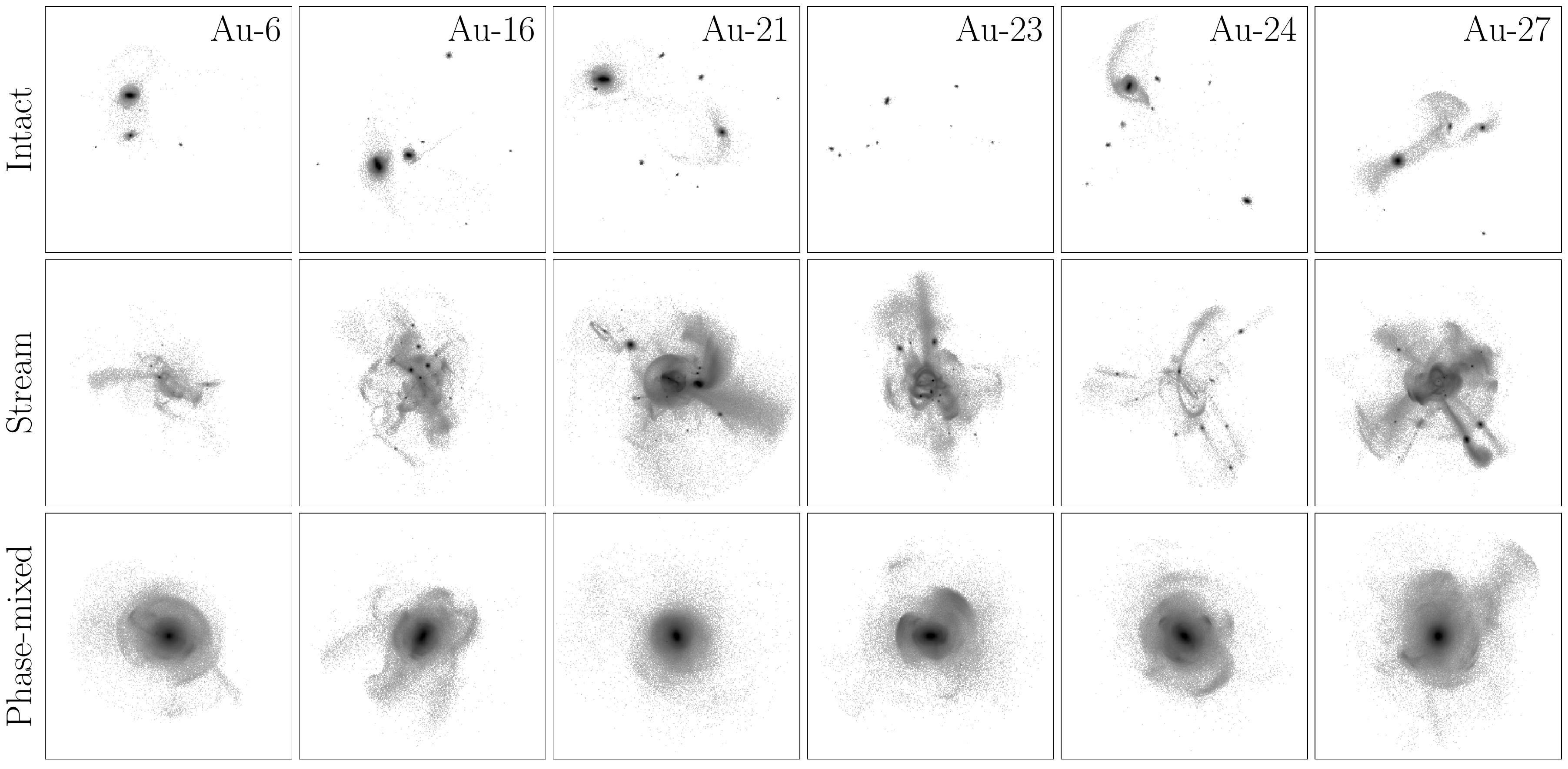}
    \caption{The accreted stellar mass of the six Auriga galaxies (Au-6, Au-16, Au-21, Au-23, Au-24, Au-27 from left to right) at level~3 resolution, with accretion events split into intact satellites (top), stellar streams (middle), and phase-mixed (bottom) morphologies.
    The shade corresponds to the amount of stellar mass in a pixel, logarithmically scaled to improve contrast in low/high-density regions.
    Note that the visible presence of tidal tails from some intact satellites, progenitors for many streams, and cold structure within phase-mixed debris reflects the continuous nature of tidal disruption.
    \textit{The Auriga sample captures a wide range of possible accretion histories for Milky Way-mass haloes, which are efficiently classified by our automated procedure (Section \ref{sec:classifying_streams}).}
    }
    \label{fig:streamportrait}
    \end{center}
\end{figure*}

This saga of nearly continuous accretion and disruption of satellites has been observed around both the Milky Way and external galaxies.
Wide-area digital sky surveys -- in particular the Sloan Digital Sky Survey \citep{York:2000}, Dark Energy Survey \citep{DES:2016}, and community DECam programs \citep{Drlica-Wagner:2021} -- have been incredibly prolific at revealing ever fainter Milky Way satellites \citep[e.g.][]{Willman:2005, Belokurov:2006booI, Bechtol:2015, Koposov:2015, Drlica-Wagner:2020, Smith:2024} and stellar streams \citep[e.g.][]{Belokurov:2006fieldofstreams, Koposov:2014, Bernard:2016, Shipp:2018}.
All-sky astrometry from the \textit{Gaia} satellite, often in combination with line-of-sight velocities and chemistry from spectrosopic surveys, has produced and characterised an abundance of nearby thin streams \citep[e.g.][]{Malhan:2018, Ibata:2019, Ibata:2021, Malhan:2021, Li:2022, Ibata:2024} and phase-mixed structures \citep[e.g.][]{Helmi:2018, Belokurov:2018, Haywood:2018, Naidu:2020, Malhan:2022pontus, Horta:2024}.
Statistical samples of satellite galaxies \citep[e.g.][]{Mao:2021, Carlsten:2021, Carlsten:2022, Mao:2024} and bright stellar streams \citep[e.g.][]{Martinez-Delgado:2010, Martinez-Delgado:2023,Miro-Carretero:2024des} have been uncovered around Milky Way-mass galaxies using deep photometric and spectrosopic data, while integrated field spectroscopic surveys have uncovered chemo-dynamical evidence of multiple accretions in inner phase-mixed stellar haloes \citep{Martig:2021, vandeSande:2024}.

In addition to encoding the assembly history of the host, disrupting satellites provide unique insights on galaxy formation at small scales and the nature of dark matter.
Tidal disruption depends on the internal structural properties of low-mass galaxies, which vary based on dark matter \citep{Du:2018, Tulin:2018} and baryonic physics \citep{Sawala:2016, Garrison-Kimmel:2019}.
Satellites can also disrupt more readily in self-interacting dark matter models due to mass loss from `ram pressure stripping' of dark matter caused by self-interactions with the host \citep{Kaplinghat:2019, Nadler:2020sidm}, resulting in clear differences in the structure of stellar haloes \citep{ForouharMoreno:2025}.
The shape and trajectory of stellar streams are extremely sensitive to the local acceleration field \citep{Johnston:2005, Bovy:2014, Gibbons:2014, Malhan:2019}, making them excellent tracers for measuring the mass and shape of the Milky Way's dark matter halo and perturbations from massive satellites \citep{Erkal:2019, Shipp:2021, Vasiliev:2021}.
Such techniques are being extended to external galaxies \citep{Pearson:2022, Nibauer:2023, Walder:2024} in preparation for upcoming deep data from facilities like Rubin \citep{Ivezic:2019}, Euclid \citep{EuclidCollaboration:2025}, and ARRAKIHS\footnote{\href{https://www.cosmos.esa.int/documents/7423467/7423486/ESA-F2-ARRAKIHS-Phase-2-PUBLIC-v0.9.2.pdf/61b363d7-2a06-1196-5c40-c85aa90c2113?t=1667557422996}{ARRAKIHS Phase 2 Proposal (public, pdf)}}.

Systems of intact satellites have been extensively studied in cosmological hydrodynamic zoom-in simulations \citep{Brooks:2014, Wetzel:2016, Sawala:2016, Garrison-Kimmel:2019} and semi-analytic models \citep{Kim:2018, Nadler:2020mwsats} in the context of resolving small-scale challenges to the $\Lambda$CDM framework \citep{Bullock:2017,Sales:2022}.
In contrast, systematic studies of disrupting satellites around Milky Way-mass haloes are relatively novel.
Studies of accreted stellar haloes that identify individual accretion events \citep{Cooper:2010, Deason:2016, Monachesi:2019, Fattahi:2020, Genina:2023} typically do not characterise the level of disruption that each accretion has experienced, instead considering the bound components of accretion events as satellites and all other accreted material as disrupted stellar halo \citep[for an analysis focused on a solar neighbourhood-like volume, see][]{Gomez:2013}.

In the pantheon of recent suites of cosmological simulations of Milky Way-mass haloes, the FIRE-2 simulations have stood out in terms of characterising disrupting satellites in detail.
\citet{Panithanpaisal:2021} identified the populations of intact satellites, stellar streams, and phase-mixed systems across 13 hosts, focusing on their origin and chemo-dynamical properties.
\citet{Shipp:2023} then performed detailed mock observations of these systems and compared their properties to the observed Miky Way satellites and streams.
They found that the number and stellar mass functions were consistent after accounting for the difficulty of detecting faint tidal tails, but that the FIRE simulations produce streams on orbits with larger pericenters and apocenters than is observed in the Milky Way.
These results raise interesting questions about the prevalence of disrupting satellites in cosmological simulations, calling for studies across a broader range of numerical prescriptions.
We answer this call by characterising disrupting satellites in the Auriga simulations, which have a large number of haloes with diverse accretion histories, were run with a different sub-grid physics model, and cover multiple resolution levels.

When moving beyond the binary distinction between satellite and stellar halo, terminology becomes important in order to capture the spatial and temporal nature of disrupting galaxies.
In addition, definitions vary across the literature\footnote{For example, `progenitor' can refer to the bound component of a disrupting system or the original bound structure prior to infall.} - we do not seek to set a standard for the field, but instead to clarify our meaning within this work.
We refer to accretion events as \textit{intact} if they have experienced little or no disruption, \textit{streams} if they have experienced tidal disruption to produce coherent structures, and \textit{phase-mixed} if they are mostly characterised as spatially smooth and occupy a phase space approaching that of the host\footnote{We note that accretions that fit this colloquial definition of `phase-mixed' are often still structured in (approximately) conserved orbital quantities long after they are smoothed in position and velocity space.}.
We refer to the full collection of accretion events, agnostic to morphological classification, as \textit{accretions}, \textit{systems}, \textit{structures}, or \textit{objects}.
If a structure still has a component that is bound to the original system, we refer to that component as the \textit{progenitor}, regardless of the overall morphology of the accretion.
A system's bound progenitor, coherent tidal tails, and/or other debris are all considered one object, to which we apply one label of intact, stream, or phase-mixed.
We sometimes refer to the collection of progenitors around a larger host as that host's \textit{satellites} or \textit{satellite system}.

In this work we characterise accretion events across the full range of disruption in the Auriga suite of cosmological simulations (Section~\ref{sec:sims}).
We separate the accreted material in Auriga into populations of intact satellites, stellar streams, and phase-mixed systems (Section~\ref{sec:stream-selection}) and match individual accretions in haloes that were re-simulated at multiple resolution levels (Section~\ref{sec:matching}).
We then analyse the contribution of each phase of disruption to the overall accreted stellar mass (Section~\ref{sec:halostructure}).
We discuss these results in the context of other studies in Section~\ref{sec:discussion} and summarise our findings in Section~\ref{sec:summary}.
This paper accompanies \citet{Shipp:2025}, hereafter Paper II, which presents the orbital properties of these systems.

\section{The Auriga Simulations}
\label{sec:sims}

We use the suite of simulated Milky Way-mass haloes from the Auriga project \citep{Grand:2017}.
These are cosmological magnetohydrodynamic zoom-in simulations of isolated haloes with halo masses\footnote{Defined to be the mass enclosed in a sphere in which the mean matter density is 200 times the critical density $\rho_\text{crit} = 3H^2(z) / 8\pi G$. Virial quantities are defined at this radius and identified with a `200c' subscript.} in the range $M_\text{200c} = 1-2 \times 10^{12}$~M$_\odot$.
They were performed with the moving-mesh code \textsc{Arepo} \citep{Springel:2010, Pakmor:2016} and a galaxy formation model that includes primordial and metal-line cooling \citep{Vogelsberger:2013}, a spatially-uniform redshift-dependent UV background for reionization \citep{Faucher-Giguere:2009}, star formation and stellar feedback \citep{Springel:2003, Vogelsberger:2013}, magnetic fields \citep{Pakmor:2017}, and black hole seeding, accretion, and feedback \citep{Springel:2005feedback, Marinacci:2014, Grand:2017}.
The Auriga simulations are described in further detail in \citet{Grand:2017, Grand:2024}.

The Auriga haloes were selected from the EAGLE dark-matter-only simulation box with comoving side length of 100 Mpc \citep{Schaye:2015}.
In addition to the narrow selection in $M_\text{200c}$, the haloes were also chosen to be relatively isolated, such that no target halo's centre was located within 9 times the $R_\text{200c}$ of any other halo that has a mass greater than 3~per~cent of the target.
The simulations used the \citet{PlanckCollaboration:2014} cosmological parameters $\Omega_\text{M} = 0.307$, $\Omega_\Lambda = 0.693$, $h = 0.6777$, $\sigma_8 = 0.8288$, and $n_s = 0.9611$.
The initial conditions for the zoom-in simulations were created using \textsc{Panphasia} \citep{Jenkins:2013}.

The original sample of 30 haloes (labelled here as Au-N, with N from 1 to 30) were simulated at `level~4' resolution\footnote{The `level' nomenclature traces back to the Aquarius \citep{Springel:2008} dark-matter-only simulations of Milky Way-mass haloes.} with baryonic element (dark matter particle) masses of 5.4(29)$\times 10^4$~M$_\odot$ and a minimum softening length of 375~pc.
Six of these haloes were re-simulated at higher `level~3' resolution (6.7(36)$\times 10^3$~M$_\odot$; 188~pc) and one halo, presented in \citet{Grand:2021}, at even higher `level~2' resolution (8.5(46)$\times 10^2$~M$_\odot$; 94~pc).
Consecutive levels differ by approximately a factor of 8 in mass resolution.
We exclude Au-1 and Au-11 from our analysis, as these haloes are experiencing massive mergers at $z=0$.
A summary of properties for the Auriga haloes is provided in Table~\ref{tab:haloes}.

\begin{table*}
\centering
\input{tables/halo-props}
    \caption{
    Summary properties of Auriga haloes analysed in this work.
    We provide the halo name and resolution level; virial mass ($M_\text{200c}$) and radius ($R_\text{200c}$) defined relative to the critical density; virial radius defined relative to the mean cosmic density ($R_\text{200m}$); total in situ ($M_\ast^\text{in situ}$) and accreted ($M_\ast^\text{acc}$) stellar mass within $R_\text{200m}$ as identified in Section \ref{sec:particle_lists}; spherical radius that encloses half of in situ stellar mass ($R_{50}^\text{in situ}$); number of intact satellites ($N_\text{intact}$), stellar streams ($N_\text{stream}$), and phase-mixed systems ($N_\text{phase-mixed}$) as classified in Section \ref{sec:classifying_streams}; and the fraction of accreted stellar mass contributed by intact satellites ($f_{\text{acc},\ast}^\text{intact}$), stellar streams ($f_{\text{acc},\ast}^\text{stream}$), and phase-mixed systems ($f_{\text{acc},\ast}^\text{phase-mixed}$).
    }
    \label{tab:haloes}
\end{table*}

Halo catalogs are constructed using the on-the-fly SUBFIND halo finder \citep{Springel:2001}.
This initially identifies haloes using a Friends-of-Friends (FOF) algorithm based on a standard linking length \citep{Davis:1985}, then separates them into gravitationally self-bound subhaloes of at least 20 particles.
The FOF step is only applied to dark matter particles (other particle types are assigned to the same group as their closest dark matter particle), while SUBFIND is run on all particle types within a FOF group simultaneously.
The merger trees were constructed in post-processing with the LHaloTree algorithm \citep{Springel:2005}, which connects FOF and SUBFIND objects both within a single snapshot and across different snapshots.
The merger trees are useful for tracking an individual object across time, including when that object merges into another.
LHaloTree identifies the descendant of a target halo by tracking which halo contains the most dark matter particles from the target in the preceding snapshot, giving higher weight to particles that were more tightly bound to the target.

It has been shown that the Auriga model produces spiral disc galaxies that are broadly consistent with a number of observations including stellar masses, sizes, and rotation curves \citep{Grand:2017}, H~\textsc{i} gas distributions \citep{Marinacci:2017}, stellar disc warps \citep{Gomez:2017}, stellar bars \citep{Fragkoudi:2020, Fragkoudi:2025}, stellar bulges \citep{Gargiulo:2019}, and magnetic fields \citep{Pakmor:2017}.
In addition, it produces systems of satellite galaxies \citep{Simpson:2018, Grand:2021} and stellar haloes \citep{Monachesi:2019} that are broadly consistent with those observed around Milky Way-mass galaxies and has been used to interpret the Milky Way's assembly history \citep{Deason:2017, Fattahi:2019}, the orbits of its satellites \citep{Riley:2019}, and to measure its mass \citep{Callingham:2019, Deason:2019}, as well as to produce realistic mock surveys of the Milky Way \citep{Grand:2018, Kizhuprakkat:2024} and M31 \citep{Thomas:2021}.
The study of \citet{Vera-Casanova:2022} is particularly relevant to our work: they characterised the brightest stellar stream around each Auriga host and connected their properties to the overall accreted stellar halo, with a focus on observations from an external perspective.

\section{Constructing the catalogue of accretions}
\label{sec:stream-selection}

In this section, we describe the procedure of identifying individual accretion events and classifying each as an intact satellite, stellar stream, or phase-mixed system.
The resulting classifications are illustrated in Figure \ref{fig:streamportrait}, which shows the accreted material of the six level~3 Auriga galaxies separated into these three categories.

\subsection{Identifying accretion events}
\label{sec:particle_lists}

We first need to unwind the accretion history of the Auriga haloes into individual accretion events and associate accreted star particles to a particular structure.
Our approach is similar to those used in prior studies on Auriga \citep{Simpson:2018, Monachesi:2019, Grand:2021} and matches how the `accreted particle lists' are constructed for the public data release \citep{Grand:2024}.

We begin by identifying every star particle within $R_\text{200m}$\footnote{This virial radius is defined relative to 200 times the mean cosmic density $\rho_\mathrm{mean}=\Omega_\mathrm{M}\rho_\mathrm{crit}$ and noted as `200m.'} \citep[$\sim$350 to $\sim$400~kpc for the Auriga haloes, $\sim$365~kpc for the Milky Way;][]{Deason:2020} of the main host at the present day.
This radial extent matches the outer radius considered in \citet{Panithanpaisal:2021} and is always larger than the 300~kpc typically considered for observational forecasts of Milky Way satellite galaxies \citep{Newton:2018, Kim:2018, Nadler:2019, Nadler:2020mwsats}.
In addition, \citet{Deason:2020} found that the edge of Milky Way-mass galaxies in Auriga and APOSTLE \citep{Sawala:2016, Fattahi:2016}, defined as a sharp drop in the logarithmic slope of the stellar density profile, can range from 0.5 to 1$\times R_\text{200m}$, with a typical value of $\sim0.6 R_\text{200m}$.
This radius adequately captures all accreted stars for our Milky Way-mass haloes, though there is a rich literature on the relationship between the `splashback' radius and a variety of halo properties \citep{Adhikari:2014, Diemer:2014, Diemer:2017, Mansfield:2017}.

Every star particle within this radius is then tracked back to the halo or subhalo in which it formed.
These (sub)haloes are connected across different snapshots using the merger trees.
Stars are assigned to the (sub)haloes that they are bound to in the snapshot immediately following their birth.
If a star was bound to a halo along the main progenitor branch of the host, it is labelled as \textit{in situ}, otherwise it is labelled as accreted.
The union of stars that form along a galaxy's main progenitor branch (except that of the main host) constitutes an individual system.
A small number of accreted stars ($\lesssim$0.3~per~cent) are not assigned to any object because they were not identified as bound to a (sub)halo at birth.

We note that in the turbulent early times of galaxy formation ($z \gtrsim 3$), it can be difficult to identify a dominant `main progenitor' of the host galaxy \citep{Horta:2024} and thus separate \textit{in situ} from accreted material, as well as individual systems from each other.
Relative to studies that explicitly ignore merger tree information earlier than a certain redshift \citep[e.g. $z=3$ in][]{Fattahi:2020}, our approach likely catalogs more individual accretion events.
Based on our automated classifications (Section \ref{sec:classifying_streams}), across the whole sample of simulations (all haloes and all levels) only 3 of the 315 accretions that cross the main host's $R_\text{200c}$ prior to $z=3$ are still streams at the present day.
Furthermore, all objects in our sample that merge into any other halo (i.e.~are no longer found by SUBFIND) prior to $z=3$ are phase-mixed and merged directly into the main host.

Throughout this work, we define an accretion event as including all stars that were born in that object.
When we quote the total stellar mass of an object ($M_\ast$), unless otherwise stated, we are referring to the sum of the present-day mass of all of its star particles, regardless of whether those particles remain bound to the progenitor at the present day.
In addition to agreeing with prior comparable analyses \citep{Panithanpaisal:2021, Shipp:2023}, this definition is motivated by efforts to infer the total stellar mass of observed streams from their measured metallicities by assuming a stellar mass-metallicity relation \citep{Li:2022}.
This total stellar mass can be larger or smaller than the `peak stellar mass' ($M_\text{peak,$\ast$}$) typically defined as the largest stellar mass bound to a halo over cosmic time.
An object that continues forming stars in its bound component as it disrupts around the host could have $M_\ast > M_\text{peak,$\ast$}$, while an object that experiences an earlier merger before falling onto the main host could have $M_\ast < M_\text{peak,$\ast$}$.
This quantity does not need to correspond to the bound stellar mass at infall, for similar reasons.

\subsection{Classifying systems}
\label{sec:classifying_streams}

We use the following criteria to determine whether the star particles associated with each accretion event identified in Section \ref{sec:particle_lists} form an intact satellite, stellar stream, or phase-mixed system at the present day:
\begin{itemize}
    \item \textit{Bound fraction} (Section~\ref{sec:fbound}). If the fraction of the object's stellar mass that is bound to the progenitor at the present day, $f_\text{bound}$, is greater than 97~per~cent, the object is labelled as intact. Otherwise:
    \item \textit{Local velocity dispersion} (Section~\ref{sec:classify-veldisp}). If the median of the local velocity dispersion of the star particles, $\sigma_\text{vel}^{50}$, is less than a stellar mass-dependent threshold, the object is labelled as a stream. Otherwise it is labelled as phase-mixed.
\end{itemize}

We only consider accretion events with at least 100 star particles to ensure that the distribution of each object is adequately captured.
This limit is more conservative than what is typically used to study satellite systems in cosmological simulations \citep[$\gtrsim$10 bound star particles;][]{Sawala:2016, Simpson:2018}, since we need to study the phase-space distribution of accretions as they disrupt.
Given the typical baryonic mass resolution at each simulation level, in addition to stellar winds that reduce the mass of star particles over time, this limit corresponds to a minimum total stellar mass of 3.7$\times 10^6$~M$_\odot$, 4.8$\times 10^5$~M$_\odot$, and 5.7$\times 10^4$~M$_\odot$ at levels 4, 3, and 2 respectively.
This also means that we effectively ignore accreted material from objects below this limit, which typically correspond to $\lesssim$0.5~per~cent of accreted stars, but can be as high as 3~per~cent for a handful of level~4 haloes.

In order to guide our classification criteria, we first select a subset of structures and classified them by eye.
We do this for all accretion events in Au-(6, 9, 10, 16, 21, 23, 24, 27) at level~4, Au-(6, 27) at level~3, and Au-6 at level~2, constituting approximately one third of the overall sample across a variety of accretion histories.
For each accretion, the lead author applied a label after looking at spatial maps of the stellar debris, similar to what is shown in Figure~\ref{fig:streamportrait}.
We refer to this sub-sample of human-classified structures as the `visually classified set'.
This set is used to calibrate the exact values of bound fraction and local velocity dispersions adopted for our automated procedure, which is then applied to all accreted systems.
The visual classifications are only used at this calibration step and occasionally for internal validation -- in all of our results, the automated classifications are used.

Even this process of visually classifying accretions further justified an automated procedure; sometimes individual accretions at different resolutions appeared remarkably similar but were assigned different labels, while assigning a label to some edge cases between stream/phase-mixed could be challenging.
Ultimately, our automated procedure classifies systems in a manner that is simple to physically interpret and maps onto three stages of disruption, as shown in Figure~\ref{fig:streamportrait}.
We discuss the shortcomings of assigning a single label to a system in Section \ref{sec:shortcomings-labelling}.

\subsubsection{Bound fraction} \label{sec:fbound}

\begin{figure}
    \includegraphics[width=1.0\linewidth]{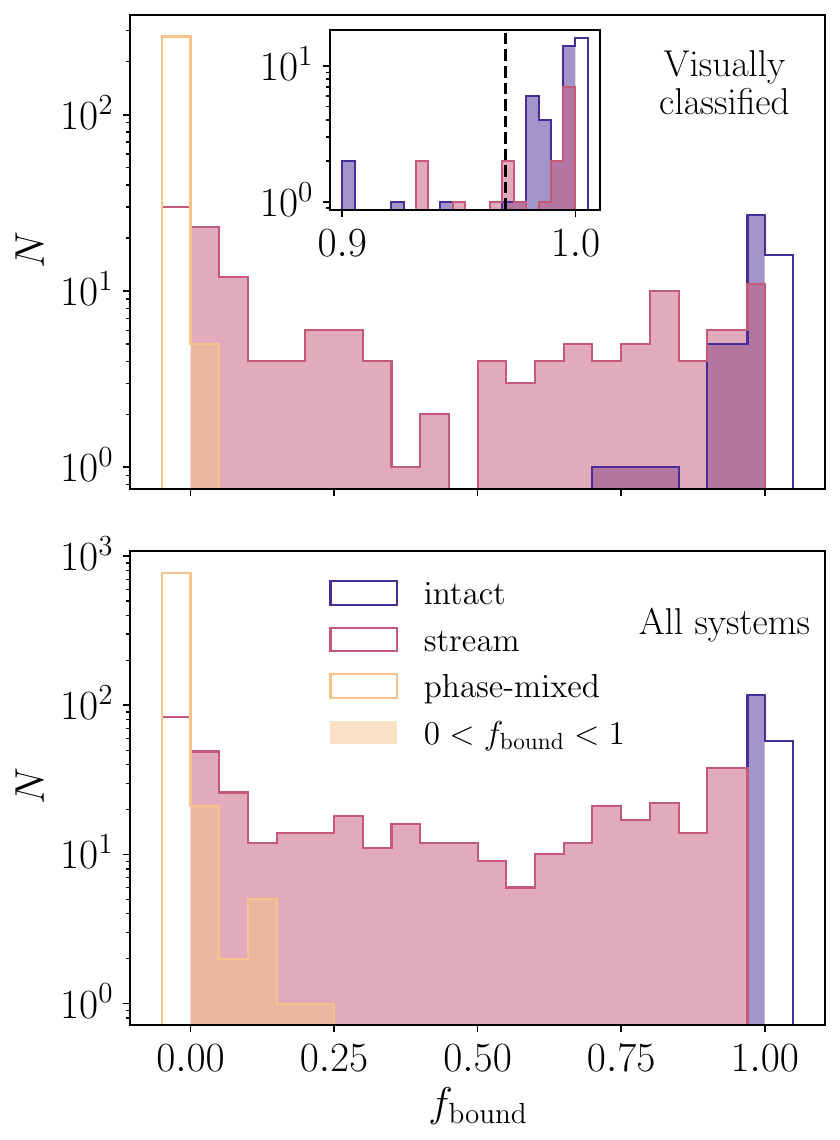}
    \caption{
    Top: fraction of stellar mass still bound to the progenitor, $f_\text{bound}$, for all accretions in the visually classified sample, split according to visual classification.
    Inset: zoom on higher $f_\text{bound}$ for the subset of objects that were classified by eye as intact satellites or streams.
    We use $f_\text{bound} = 0.97$ (dashed vertical line) as the boundary between intact satellite and stream for our automated classification.
    Bottom: distribution of $f_\text{bound}$ for all accretions, split by classification assigned automatically.
    Shaded regions indicate objects that have $0 < f_\text{bound} < 1$, which combined with the right-most bin of $f_\text{bound} = 1$ systems would traditionally be treated as `satellites' in simulation studies.
    \textit{The majority of satellites with a bound progenitor have lost substantial amounts of stellar mass.}
    }
    \label{fig:fbound-dist}
\end{figure}

The distinction between intact satellites and stellar streams is based on how much stellar mass the progenitor has lost by the present day.
The fraction of the object's stellar mass that is bound to the progenitor is determined by SUBFIND and only includes stellar mass formed in that object (i.e.~excluding other mergers whose debris is now bound to that SUBFIND subhalo).
The top panel of Figure~\ref{fig:fbound-dist} shows the distribution of $f_\text{bound}$ for our visually classified sample.
Nearly all intact satellites have $f_\text{bound} \sim 1$ (across all haloes and resolution levels, 58 objects have $f_\text{bound} = 1$ and 84 have $0.99 < f_\text{bound} < 1$), while the few instances that are substantially below this are typically low-mass systems that have only lost a few star particles (out of 100-200 total).
We set the threshold at $f_\text{bound} = 0.97$, which captures nearly all structures that were visually classified as intact (inset panel of Figure~\ref{fig:fbound-dist}).
In addition, when sorting the full sample by descending $f_\text{bound}$, the transition from where most systems were classified as intact (with a few stream interlopers) to where most systems were classified as streams (with a few intact interlopers) also occurs near $f_\text{bound} = 0.97$.

The bottom panel of Figure \ref{fig:fbound-dist} shows the distribution of $f_\text{bound}$ for all accretion events in our sample, now separated according to our automated classification criteria.
The overall distribution is similar to the visually classified sample, with two sharp peaks at $f_\text{bound} = 0$ and $f_\text{bound} = 1$.
Overall, a picture emerges where most systems with a bound progenitor (ie.~$f_\text{bound} > 0$), which would be traditionally labelled as `satellites' in simulation studies, have lost substantial amounts of stellar mass.
Of all objects with a bound progenitor at the present day, 67~per~cent have lost enough stellar mass to no longer be considered intact satellites.
Even if our threshold definition of $f_\text{bound} = 0.97$ may seem arbitrarily high, lowering it does not affect this qualitative result -- 53 (40)~per~cent of objects with a bound progenitor have $f_\text{bound} < 0.8$~(0.5).
We discuss the implications of this result in Section \ref{sec:discussion}.

\subsubsection{Local velocity dispersion} \label{sec:classify-veldisp}

\begin{figure}
    \includegraphics[width=1.0\linewidth]{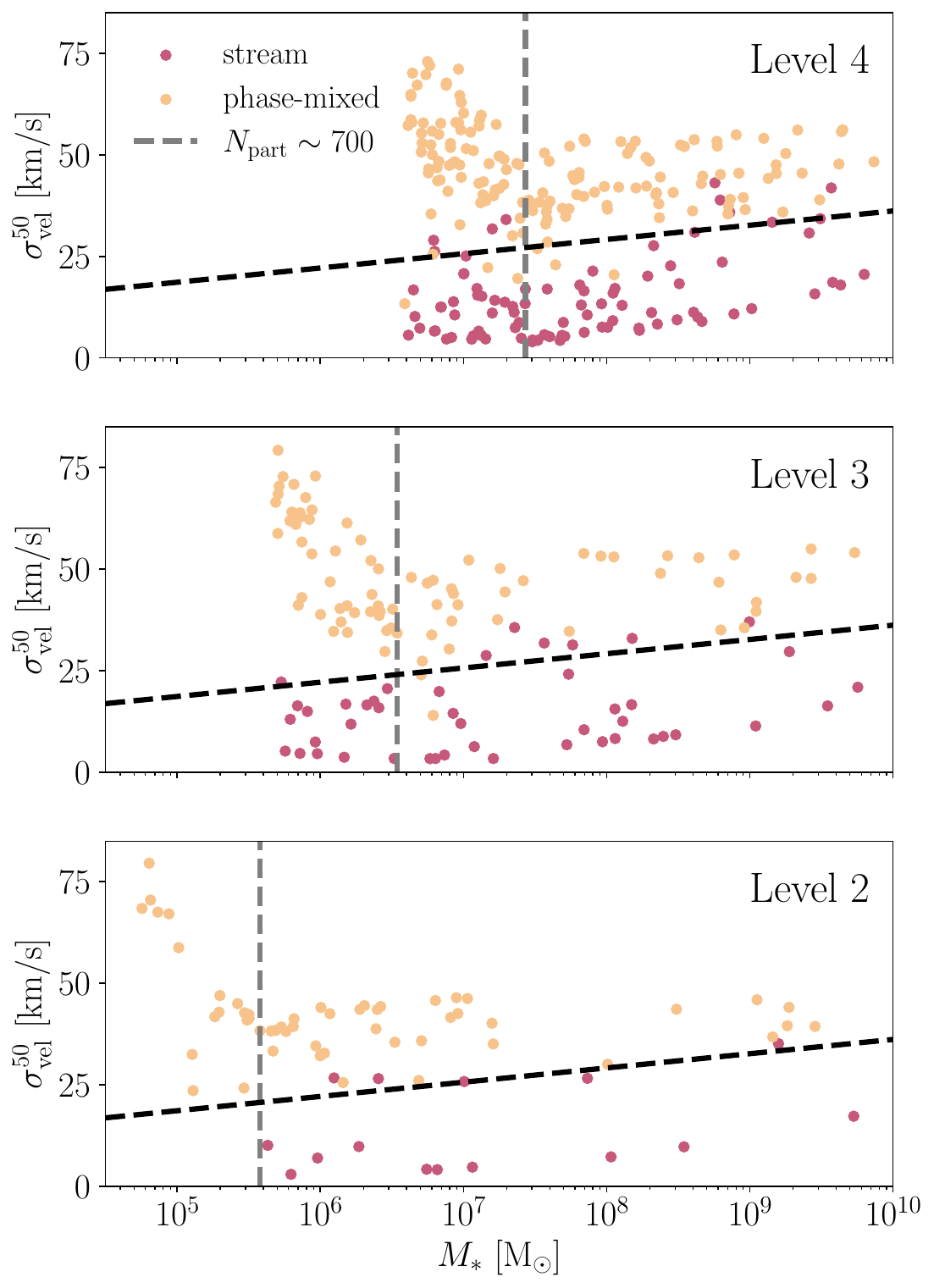}
    \caption{
    Stellar mass-velocity dispersion plane used to classify objects as streams or phase-mixed.
    Points are coloured based on visual classifications, while the black dashed line corresponds to the plane that optimally separates the two populations (Equation \ref{eqn:local_vel_disp}) we adopt for our automated procedure.
    The vertical dashed line corresponds to the stellar mass of the most massive object with fewer than 700 star particles, approximating the transition from using 1~per~cent of particles to the required minimum of 7 neighbours in the local velocity disperion calculation.
    \textit{At fixed stellar mass, streams have colder local velocity dispersions than phase-mixed debris, which we use to automatically classify the full sample of accretions.}
    }
    \label{fig:veldisp-smass}
\end{figure}

As disrupting satellites continue to evolve, they eventually become phase-mixed systems that are spatially smooth.
In our visually classified set, stellar streams have colder velocity distributions than phase-mixed systems of similar stellar mass.
We base our stream/phase-mixed distinction on this fact, specifically considering the median of the distribution of local velocity dispersions for all star particles in a given accretion.

To compute the local velocity dispersion for an individual star particle, we find its nearest neighbours in phase space among star particles associated with the same accretion event.
We consider both position and velocity information when identifying neighbours, which avoids assigning neighbours that are close in position but not in velocity (e.g.~if a stream has multiple wraps).
However, this introduces a relatively arbitrary choice of defining a metric that combines the two.
Taking inspiration from studies of phase-space clustering in the stellar halo \citep[e.g.][]{Cooper:2011}, we adopt the distance metric between two particles (denoted $i$ and $j$)
\begin{equation}
    \Delta^2_{ij} = |\boldsymbol{x}_i - \boldsymbol{x}_j|^2 + w_v^2 |\boldsymbol{v}_i - \boldsymbol{v}_j|^2
\end{equation}
which scales the particles' velocities with a weighting factor $w_\text{v}$ that has units of kpc~km$^{-1}$~s, such that $\Delta$ has units of kpc.
The local velocity dispersion is then the velocity dispersion among the $k$ neighbours with the smallest $\Delta$, where $k$ is taken to be 1~per~cent of the total number of star particles in the object or a minimum of 7 particles, whichever is larger.
We repeat this exercise to compute the local velocity dispersion for all star particles in the same accretion event, then take the median of this distribution for our classification purposes (see below).
Fixing the fraction of particles in the accretion, rather than the number of neighbours $k$, controls for the increased number of particles (and therefore smaller amount of phase space occupied by each particle) for the same structure at higher resolution levels.
The absolute minimum of 7 particles prevents a biased calculation of the local velocity dispersion based on too few neighbours and is the same absolute minimum adopted in \citet{Panithanpaisal:2021}.
The adopted value of 1~per~cent is relatively arbitrary -- larger values push the definition of `local' further away in phase space from the particle of interest, while smaller values result in more objects falling back to the minimum of 7.
Across the full sample, $\sim$55~per~cent of objects are under the 1~per~cent regime (i.e.~have more than 700 star particles).

Figure \ref{fig:veldisp-smass} shows the median local velocity dispersion against stellar mass for accretions in the visually classified sample.
We find that at fixed stellar mass, objects that were classified as phase-mixed have preferentially larger median local velocity dispersions than objects that were classified as streams.
The transition from stream to phase-mixed occurs in roughly the same space across different resolution levels, suggesting that the physical processes that drive phase-mixing are not impacted by the lower resolution of Auriga level~4.
The uptick in velocity dispersion for phase-mixed objects at low stellar mass occurs near a total particle count of 700, where the $k$ adopted for calculating the local velocity dispersion changes from being 1~per~cent of particles in the object to the minimum of 7.

We use a linear support vector machine\footnote{Implemented in scikit-learn \citep{scikit-learn}.} \citep[SVM;][]{Cortes:1995} to determine the line that maximally separates the two populations in the visually classified set
\begin{equation}
    \label{eqn:local_vel_disp}
    \sigma^{50}_\text{vel} = A \log \left( \frac{M_\ast}{M_\odot} \right) + B,
\end{equation}
where $\sigma^{50}_\text{vel}$ is the median of the local velocity dispersion and $M_\ast$ is the total stellar mass.
We run this SVM step on the full visually classified set of streams and phase-mixed systems, including data at all three resolution levels.
In our automatic classification of all systems, accretions with $f_\text{bound} < 0.97$ that fall above this line are classified as phase-mixed, while those below the line are classified as streams.
Intact satellites nearly always have the lowest $\sigma^{50}_\text{vel}$ for their stellar mass.
There is a modest positive slope on the mass dependence of Equation \ref{eqn:local_vel_disp}, as well as a minimum possible median local velocity dispersion that grows with stellar mass, which reflects the fact that more massive accretions have higher velocity dispersions prior to infall.

A number of the `hyperparameters' in this step ($w_v$, $A$, $B$) were calibrated via trial-and-error to produce an automated classification that is reasonably representative of the visually classified set.
In particular, the choice of $w_\text{v}$ is not obvious a priori.
Studies on phase-space clustering in the stellar halo have adopted values based on the observed span of a particular survey \citep{Starkenburg:2009, Xue:2011, Janesh:2016, Yang:2019substructure} or calibrated to place a turnover in the separate spatial and velocity correlation functions at the same scale \citep{Cooper:2011}.
After re-running our full analysis with several $w_v$ values, we found that many structures had the same classification after (re-)calibrating the stream/phase-mixed separation line, even though the individual values of $\sigma^{50}_\text{vel}$ varied considerably\footnote{As an illustrative example, for extremely large values of $w_v$, $\Delta$ becomes dominated by separation in velocity and the resulting velocity dispersion of the `nearest neighbours' will be lower for nearly all objects.}.
We found that $w_v = 4$~kpc~km$^{-1}$~s provides the optimal results for a few particular configurations of accretions; lower values of $w_v$ tend to result in phase-mixed systems that sit very deep in the host's potential being mis-classified as streams, while higher values of $w_v$ can improperly apply the phase-mixed label to structures with nearly all of their stellar mass in tidal tails embedded in a diffuse envelope.
The resulting parameters for the stream/phase-mixed separation plane in Equation~\ref{eqn:local_vel_disp} are $(A,B)=(3.51,1.08)$~km~s$^{-1}$, shown as a dashed line in Figure~\ref{fig:veldisp-smass}.

We caution against comparing our computed $\sigma_\text{vel}^{50}$ values, which we only use to establish a relative difference between cold streams and phase-mixed debris for classification purposes, to observed velocity dispersions.
This is both because we consider the full set of star particles (e.g.~progenitor and tidal debris) in our calculation and because our calculation is sensitive to the adopted $w_v$.
Analyses that seek to make this comparison should begin with the star particles in a given accretion event and compute the desired quantity directly.

\subsubsection{Comparison to \citet{Panithanpaisal:2021}} \label{sec:nondh-comparison}

Our approach is heavily inspired by that used in \citet{Panithanpaisal:2021} to classify accretion events in the FIRE suite.
Their method differs from ours in the following ways:
\begin{enumerate}
  \item They only consider objects with a number of star particles between 120 and $10^5$.
  \item Rather than base the intact/stream boundary on $f_\text{bound}$, they use a boundary based on how spatially compact the object is. Specifically, they require at least one pair of star particles to be separated by at least 120 kpc to be classified as a stream.
  \item When computing the local velocity dispersion, they set the number of nearest neighbours $k = 20$ for objects with more than 300 star particles and $k = 7$ otherwise.
  \item When computing the local velocity dispersion, they vary the velocity-to-distance scaling $w_v$ on an accretion-by-accretion basis.
  They use $w_v^i = \sigma^i_\text{pos} / \sigma^i_\text{vel}$, where $\sigma_\text{pos}$ and $\sigma_\text{vel}$ are the position and velocity dispersions of all star particles in each accretion $i$.
\end{enumerate}

The changes that we made accommodate analysing data at multiple resolution levels (iii), are more convenient to calculate and do not exclude certain orbital configurations (ii), or are implementation choices that do not ultimately affect results or interpretation (i and iv).
In Appendix \ref{app:fire-classification} we show the results of applying the same selection criteria to the level~4 Auriga data, keeping only our treatment of selecting $k$ since this sample has lower mass resolution than the FIRE haloes analyzed in \citet{Panithanpaisal:2021}.
In short, we find it unlikely that any differences between disrupting systems in FIRE and Auriga are due to definitions or classification methodology.

Finally, we caution that our exact classification algorithm (the values of $w_v$, $A$, $B$, $k$ listed above) is not intended to be plug-and-play for analysis of other simulation suites.
For example, the $f_\text{bound} = 0.97$ intact/stream separation determined in Section \ref{sec:fbound} could have arisen from particular minutae of the Auriga model or SUBFIND halo finder.
However, the general approach (a classification based on boundaries in $f_\text{bound}$ and $M_\ast$ vs. $\sigma_\text{vel}^{50}$) should be broadly applicable, and the modifications we have made to the method outlined in \citet{Panithanpaisal:2021} are likely to be useful for any analysis that considers simulations at different mass resolutions.

\section{Matching across resolutions}
\label{sec:matching}

\begin{figure}
    \includegraphics[width=1.0\linewidth]{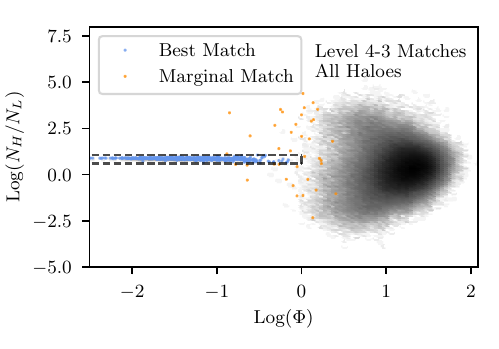}
    \caption{
    Illustration of the method used to match accretions across different resolution levels.
    The ratio of particle numbers $N_H / N_L$ and the metric $\Phi$ (as defined in Equation \ref{eq:match_metric}) are computed for each possible pairing of luminous accretions between level~4 to 3.  An intensity map of their distribution is shown in gray.
    Pairs that fall within the dashed box are considered matches (blue points); if a low resolution object has no match then the pairing that minimizes $\Phi$ is identified as a marginal match (yellow points).
    \textit{Accretion events are efficiently matched to their high-resolution counterparts using a combination of $N_H / N_L$ and $\Phi$.}
    }
    \label{fig:matching-metrics}
\end{figure}

Cross-matching accretion events between simulations with the same cosmological perturbations but different resolutions requires an algorithm to match Lagrangian volumes that collapse to form the same structures.
Our strategy for matching the accretions across resolutions uses the dark matter particles associated with star particle groups in position space in the initial conditions.

Our procedure is as follows.
We first identify star particles that were born in the same merger tree branch (Section~\ref{sec:particle_lists}).
We then identify the simulation snapshot along the merger tree branch where the total mass (dark matter and stars) of that object peaks and capture its associated dark matter particles according to the \textsc{FOF} and \textsc{SUBFIND} algorithms at this snapshot.
We track those dark matter particles back to the initial conditions and compute their average position in all three dimensions $\boldsymbol{\mu}=(\mu_x, \mu_y, \mu_z)$ together with the standard deviation in their positions $\boldsymbol{\sigma} = (\sigma_x, \sigma_y, \sigma_z)$.

We then compare the debris groups (dark matter particles associated to accretion events in initial condition space) from two simulations for the same halo separated by one resolution level.
We compute a comparison metric $\Phi$ for each high resolution ($H$) and low resolution ($L$) debris group pair that is the magnitude of a vector with these components for each direction $i$:
\begin{equation} \label{eq:match_metric}
    \Phi_i = \frac{\mu_{i,L}-\mu_{i,H}}{\sqrt{\sigma^2_{i,L} + \sigma^2_{i,H}}}.
\end{equation}

We also consider the ratio in dark matter particle number between each debris group pair.
The difference in mass resolution between each resolution level is a factor of 8.
Matched structures that are converged in dark matter mass therefore should have a particle ratio of approximately 8.
Figure \ref{fig:matching-metrics} shows $\Phi$ and the particle ratios for all possible matched debris group pairs for all halos simulated at both levels~3 and 4.
We can see that at small values of $\Phi$ there is a concentration of pairs with a particle ratio of 8; these indicate highly significant matched pairs.

To assign a best match for each debris group in the low resolution simulation, we find the pairing with the high resolution simulation that has the smallest value of $\Phi$.
If this pairing has a value $\Phi < 1$ and the particle ratio is within a factor of 2 of 8 (dashed box in \ref{fig:matching-metrics}) the group is assigned to be a `match'.
In cases where the minimum $\Phi$ corresponding group lies outside this region, the group is assigned to be a `marginal match'.
If multiple low-resolution groups are marginally matched to the same high-resolution group, we label the most massive low-resolution object as the marginal match and all other objects are labelled `unmatched'.
There are no cases where multiple low-resolution groups are high-quality matches to the same high-resolution group.

\begin{figure*}
    \includegraphics[width=1.0\linewidth]{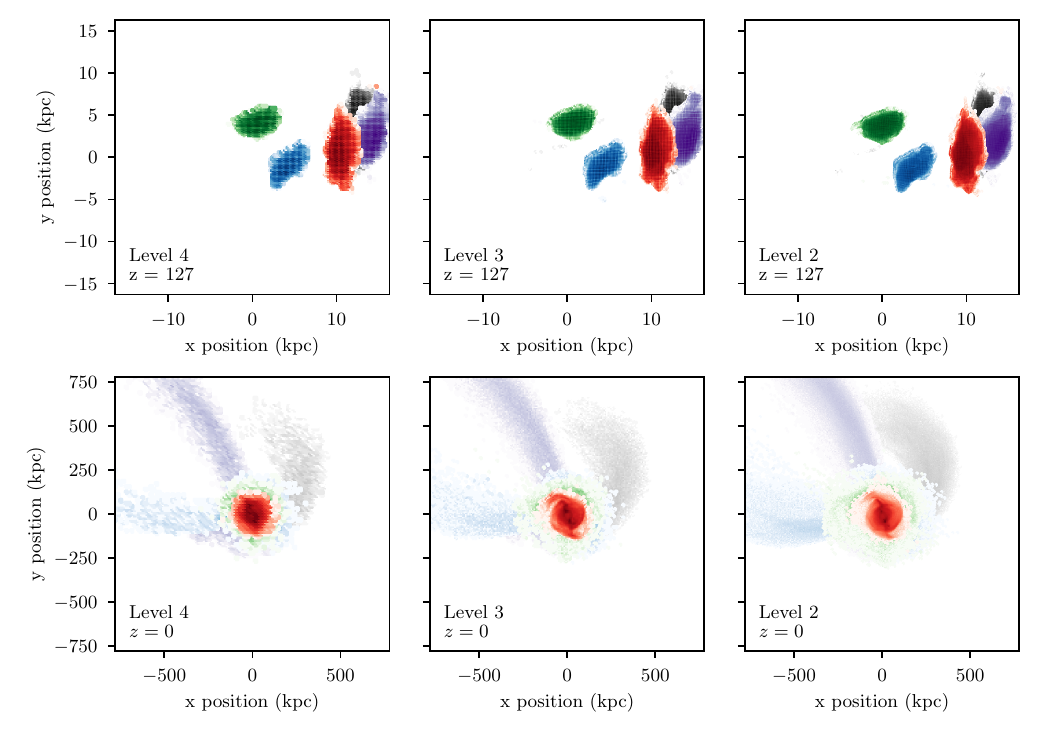}
    \caption{Demonstration of the matching procedure for merged systems across all 3 resolution levels in Au-6.  The five systems shown are the five systems categorized as streams in the level~4 simulation of this halo.  An intensity map of the dark matter mass density is shown in a matching color between the ICs ($z=127$) and the final snapshot ($z=0$) for systems matched across resolutions. The images are centred on the most bound particle for the main host at the present day and positions are plotted in physical coordinates. Particles in the initial conditions are laid out on a grid, which causes the grid like appearance in the IC projection panels. \textit{Our algorithm for matching structures between different resolution simulations produces matches with a high degree of physical coherence at both the beginning and end of the simulation.}}
    \label{fig:dm_matching_combo}
\end{figure*}

This procedure produces matches with a high degree of physical coherence in dark matter distributions over the course of the simulation and across different resolution levels.
Figure \ref{fig:dm_matching_combo} shows the dark matter particle positions for matched structures categorized as stellar streams by our classification procedure.
These systems are consistent in their spatial distributions in the initial conditions (used for matching) but also in the final snapshot at the present day.

Matches are only considered between luminous objects that cross $R_\text{200m}$ by $z=0$.
We only seek matches for each system in the lower resolution simulation among systems in the corresponding higher resolution simulation.
We do not seek matches in the opposite direction, because it is expected that some systems in the higher resolution simulation at lower masses may not have a match at lower resolution due to resolution effects that suppress star star formation (e.g. from under-resolved gravitational potentials that are not steep enough to allow gas to cool and collapse).
In addition, we exclude objects with fewer than 100 star particles when constructing our catalogue of accretion events, so low-mass objects that form at both low and high resolution may not be included if they are under this threshold in the low resolution run.

A small number of systems in the low resolution simulation do not find a match.
There are some reasons this can arise because of the way we have constructed our sample of debris groups for comparison.
For example, if a system is on first infall at the present day, it is possible that the corresponding system one resolution level higher has not yet crossed $R_\text{200m}$ and is therefore left out of our sample of comparison systems.

We considered several matching techniques and initially tried the method described in \citet{Grand:2021}, which finds the high resolution counterpart that minimizes the sum of positional offsets across all snapshots prior to infall.
However, that method was applied to simulations that had the same number of snapshot outputs.
Five of our level~3 resolution simulations have only 64 snapshots, instead of the 128 snapshots for levels 4 and 2.
We found the method of \citet{Grand:2021} did not produce good matches in these cases.
We also considered matching dark matter particles to their nearest neighbour in initial conditions and comparing debris groups on a particle-by-particle basis.
However, we found that this more complex technique did not give better matches (as determined by visual comparison of dark matter distributions at $z=0$ and host-centric distance vs.~time) than our adopted approach.

Across the six halos with both level~4 and level~3 resolution runs, 27 intact satellites, 67 streams, and 108 phase-mixed systems present in level~4 were matched by our method.
The number of systems in the three categories that were not matched (either `marginal' or `unmatched') the level~4 runs were 4, 1, and 10 respectively, a match rate of over 90~per~cent for each morphology across all 6 halos.
There were systems present in the level~3 simulations that were not matched to a level~4 systems, but this is expected due to resolution effects of the simulations \citep{Grand:2021}.
For the level~2 run of Au-6, all intact satellites and streams that were present in the level~3 simulation were present in the level~2 simulation.
Only 1 phase-mixed system in the level~3 simulation was not matched in the level~2 simulation.

\begin{figure*}
    \includegraphics[width=1.0\linewidth]{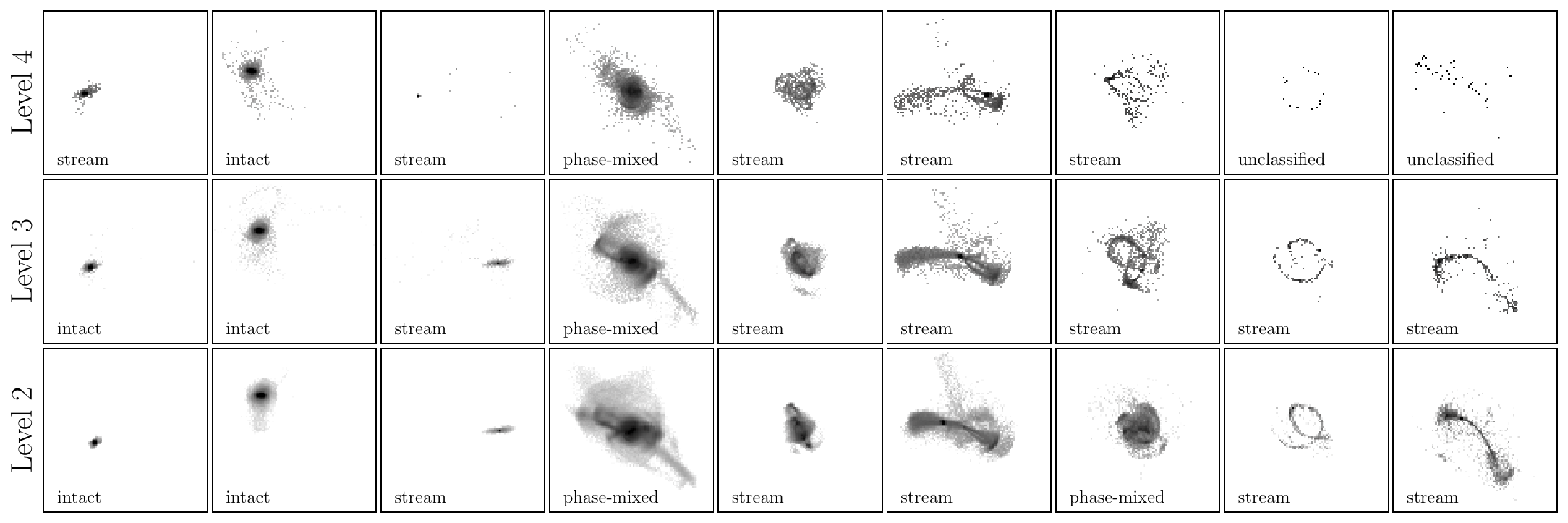}
    \caption{
    Stellar morphologies of individual structures in Au-6 matched across three resolution levels (increasing in resolution from top to bottom).
    We provide the classification label from Section \ref{sec:classifying_streams}; objects labelled `unclassified' have fewer than 100 star particles.
    The shade corresponds to the amount of stellar mass in a pixel, logarithmically scaled to improve contrast in low/high-density regions (the normalisation is consistent across resolution but not for different columns, as the objects span a wide range in stellar mass).
    Cases where the morphology does not match across all levels (third from the left, third from the right) correspond to the same accretion event experiencing slightly different accretion timing (see Paper II for further detail).
    \textit{The stellar morphologies of accretions in Auriga are remarkably converged across resolution, often localizing the progenitor to a similar phase along its orbit.}
    }
    \label{fig:matching-stellar-maps}
\end{figure*}

\begin{figure*}
    \includegraphics[width=1.0\linewidth]{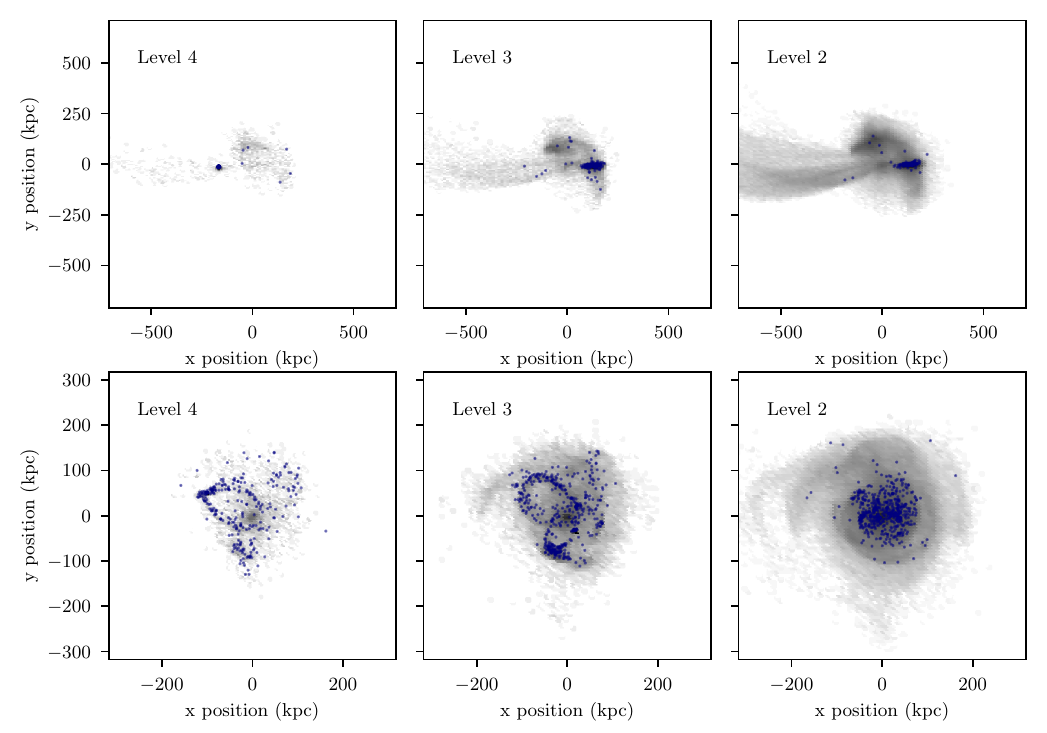}
    \caption{A detailed view of two systems across resolutions that have apparent differences in their stellar distributions from Figure~\ref{fig:matching-stellar-maps}.  These two systems are shown in stellar mass in Figure~\ref{fig:matching-stellar-maps} (the 3rd and 7th columns from that figure); here, we show both in their stellar component (blue points) and their dark matter distributions (gray intensity map).  The star particles in the level~2 panels were under-sampled by a factor of 8 for visual clarity.  Their underlying dark matter distributions have a high degree of spatial coherence across resolutions.  The dark matter of these systems is also shown in the initial conditions in Figure \ref{fig:dm_matching_combo} (the blue and green heatmaps for these two cases).  There we see that the dark matter occupies the same Lagrangian volume at high redshift.  \textit{These examples demonstrate that differences in the stellar light distributions between resolutions can in some cases belie the total mass distribution and indicate a differing dynamical history rather than a mismatch.}}
    \label{fig:dm_mismatches}
\end{figure*}

Figure~\ref{fig:matching-stellar-maps} shows the distribution of stellar mass for individual objects matched at the three resolution levels for Au-6.
These matches show a high degree of agreement in spatial distributions, though there are some cases of mismatch in final stellar positions.
We investigate these cases further in Figure~\ref{fig:dm_mismatches}, and find that their dark matter distributions that match more closely, showing that these are in fact the same accretion event.
The different stellar morphologies arise from slightly different accretion times and number of experienced pericentric passages (see Paper II for further detail).

\section{The three-phase structure of accreted stellar mass} \label{sec:halostructure}

After taking our steps to identify and characterise accretion events (Section \ref{sec:stream-selection}) and pair low resolution systems with high resolution counterparts (Section \ref{sec:matching}), we have effectively assembled a catalogue of accretion events (complete with useful meta-data for each accretion, like the fraction of stellar mass still bound to the progenitor) that can be paired with the simulation particle data for any number of interesting science applications.
As a demonstrative example, in this section we examine the contribution of our three morphological classifications (intact/stream/phase-mixed) to the overall accreted material in Auriga.
We note that for the following section, we assign the total stellar mass of an object to that object's morphological class (e.g.~a system classified as a stream with $f_\text{bound} = 0.5$ and distinct tidal tails would have all of its star particles contributing to the `stream' category), consistent with our definition in Section~\ref{sec:particle_lists}.

\begin{figure*}
    \includegraphics[width=1.0\linewidth]{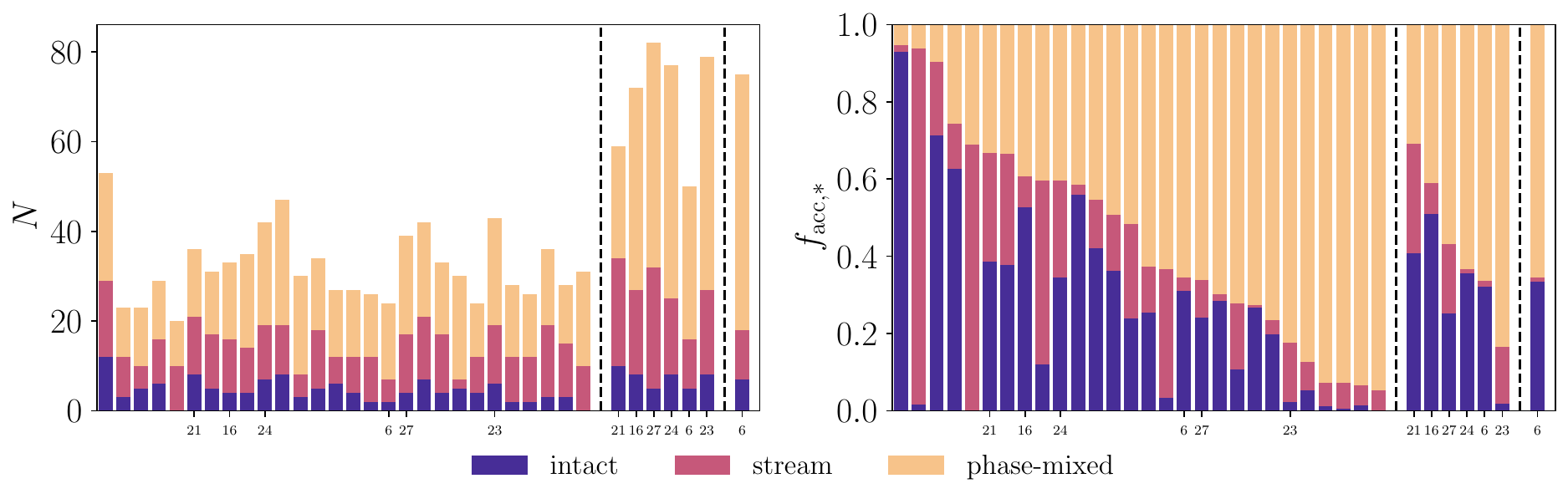}
    \caption{Left: number of accretion events in each halo, split by classification.
    Right: fraction of accreted stellar mass in each classification.
    Haloes are ordered within a given level according to the fraction of accreted stellar mass that is in a phase-mixed system; ordering is consistent between left and right panels.
    Information for specific haloes is listed in Table~\ref{tab:haloes}.
    Vertical dashed lines separate resolution levels, with 28, 6, and 1 halo(es) in levels 4, 3, and 2, respectively.
    Haloes that are re-simulated at higher resolution are specified to assist comparison.
    \textit{While the number of systems in each category is roughly consistent across hosts, the fraction of stellar mass contributed by each phase of disruption spans a wide range of possibilities.}
    }
    \label{fig:barcharts}
\end{figure*}

Figure \ref{fig:barcharts} shows the overall number of accretions (left panel) and fraction of accreted stellar mass within $R_\text{200m}$ (right panel) contributed to each halo from the different morphological classes (these values are also listed individually in Table \ref{tab:haloes}).
Each column corresponds to a simulated halo, with the columns sorted in both panels by the fraction of stellar mass contributed by phase-mixed systems.
Different resolution levels are separated by dashed lines.
At level~4 resolution, there are typically 5-10 intact satellites, 10-15 stellar streams, and 15-20 phase-mixed systems.
The relative ratios are roughly the same for the level~3 haloes, with more total accretions per halo since these simulations resolve the formation of lower mass galaxies.

There is considerably more variety across haloes at a given level when we instead consider the fraction of accreted stellar mass contributed by each morphological class.
The level~4 haloes span nearly the full range of possible values for fraction of stellar mass contributed from phase-mixed objects, from $\sim$5 to $\sim$95~per~cent.
The relative contribution of intact satellites and streams to the `cold' component of the accreted stellar mass varies wildly; often this is driven by the state of disruption of the most massive satellite.
These results are qualitatively consistent with level~3, whose haloes span a similarly large range of possible phase-mixed contributions, $\sim$30 to $\sim$80~per~cent, that encompasses where the bulk of level~4 haloes reside.

From Figure~\ref{fig:barcharts} and Table \ref{tab:haloes}, it is also clear that Au-6, the halo simulated at all three resolution levels, has a very quiet merger history.
It has so few individual accretion events that the number of objects in the level~2 run is similar to other haloes at level~3, even though level~2 resolves nearly an order of magnitude smaller objects in terms of stellar mass.
In addition, very few of the accretion events that Au-6 experiences are stellar streams at the present day.
Among its peers in levels 4 and 3, Au-6 always has one of the smallest collections of streams by number, and at all resolution levels its fraction of stellar mass in streams is below 4~per~cent.
For this reason, we focus only on levels 4 and 3 for the rest of this section.

\begin{figure*}
    \includegraphics[width=\textwidth]{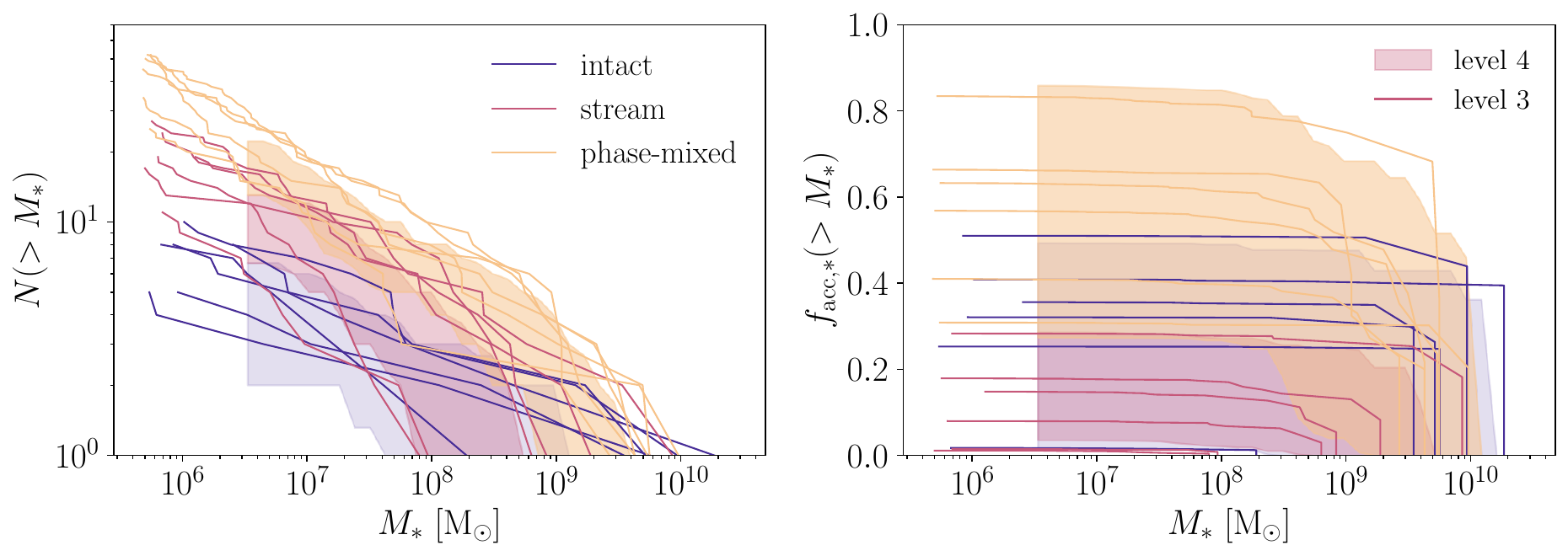}
    \caption{
        The build-up of accreted stellar mass in Auriga, split by morphological classification.
        Left: number of accretions above a given stellar mass.
        Right: fraction of accreted stellar mass contributed by objects above a certain stellar mass.
        The shaded regions indicate the 16th and 84th percentiles for the level~4 haloes, while the individual lines correspond to individual level~3 haloes.
        \textit{While stellar streams tend to outnumber intact satellite galaxies by a factor of two, they typically contribute least to the overall accreted stellar mass.
        At all phases of disruption, a few massive accretion events contribute the bulk of stellar mass.}
    }
    \label{fig:halo-structure}
  \end{figure*}

Rather than restrict ourselves to overall contribution of each morphology to the accreted stellar mass, we can also examine their contributions as a function of stellar mass of the individual accretions.
The left panel of Figure \ref{fig:halo-structure} shows the cumulative number of accretion events of each morphological type as a function of total stellar mass (as defined in Section \ref{sec:particle_lists}), in the style typically presented for the luminosity function of satellite galaxies.
At the most massive end ($M_\ast \gtrsim 10^{9.5}$) accretions tend to either be intact satellites or phase-mixed, depending on the accretion time onto the main host and reflecting the impact of dynamical friction reducing the amount of time high-mass objects remain dynamically cold.
Below this mass, there tend to be more streams than intact satellites and more phase-mixed systems than streams, by roughly a factor of 2 in both cases.
These trends are consistent across resolution as well, with level~3 haloes (represented as individual lines) largely within the scatter of the level~4 sample (shaded regions), though as a group the level~3 haloes tend to have more accretion events than the full sample of level~4 haloes.
Some of the level~3 haloes (Au-21, Au-23, Au-27) were selected for re-simulation based on massive satellite interactions \citep{Grand:2018}, which helps explain the apparent disagreement between levels 3 and 4 for intact satellites on the massive end.
We discuss the effect of halo selection and halo-to-halo scatter further in Section \ref{sec:auriga-issues}.

The right panel of Figure \ref{fig:halo-structure} shows the same information, but re-weighted by contribution to the total accreted stellar mass.
Here we see that while there are more stellar streams than intact satellites, streams tend to contribute less accreted stellar mass.
While there is substantial overlap in the different shaded regions corresponding to level~4 haloes (reflecting the wide range of possibilities also seen in the right panel of Figure \ref{fig:barcharts}), phase-mixed systems tend to contribute the most accreted stellar mass, then intact satellites, then stellar streams.
These results are consistent with the level~3 sample.

In Figure \ref{fig:halo-structure} we also recover the familiar result that most of the accreted stellar mass comes from one (or a few) massive accretion(s).
This is a natural outcome of the halo mass function set by $\Lambda$CDM and the steepness of the stellar mass-halo mass relation that has been validated by both semi-analytic modeling \citep{Bullock:2005, Cooper:2010, Deason:2016, Amorisco:2017} and hydrodynamic simulations \citep{Monachesi:2019, Santistevan:2020, Fattahi:2020, Wright:2024}.
Perhaps unsurprisingly, the right panel of Figure \ref{fig:halo-structure} demonstrates that this picture holds across the three morphological types, with the overall contribution from intact, stream, and phase-mixed morphologies each dominated by a few massive objects.
A related effect is that the contribution of a morphological type to the overall halo is largely determined by the most massive objects of that morphology, which tend to be either very recent intact accretions or phase-mixed at the present day.

\begin{figure}
    \includegraphics[width=\linewidth]{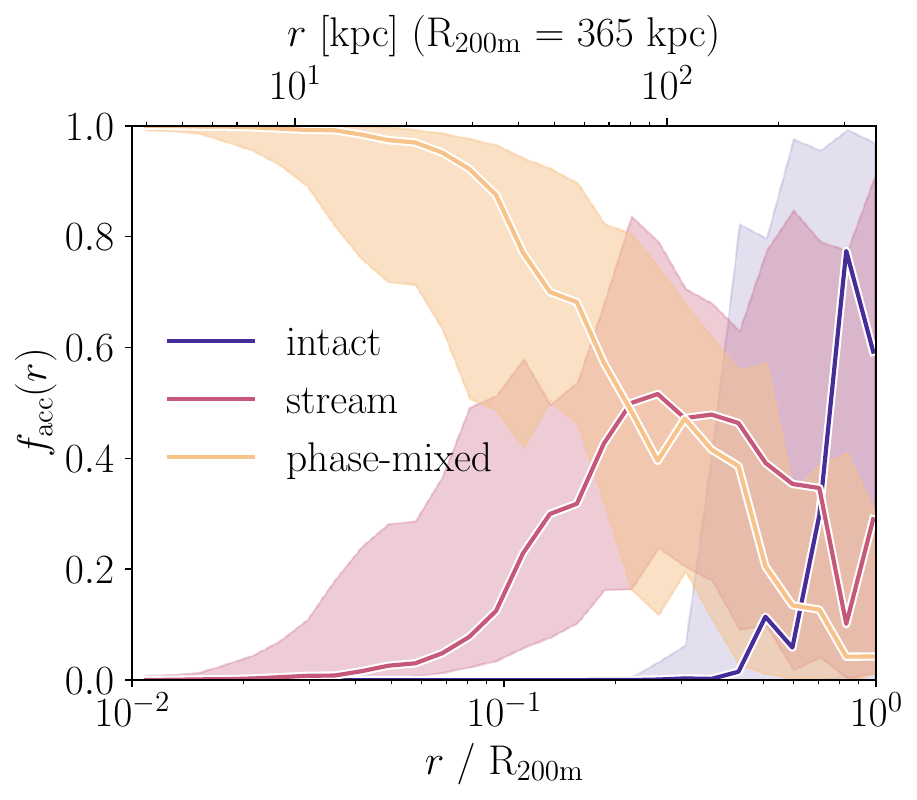}
    \caption{
        The fraction of accreted stellar mass contributed by each morphology class at a given radius (normalised to $R_\text{200m}$).
        Solid lines indicate the median and shaded regions the 16th and 84th percentiles for the level~4 haloes.
        While we only show the level~4 result for visual clarity, the distributions for levels 3 and 2 are consistent.
        On top axis we show this in physical distance, adopting the estimate of the Milky Way's $R_\text{200m}$ from \citet{Deason:2020}.
        \textit{The inner region of accreted stellar mass is dominated by phase-mixed systems, giving way to stellar streams near 0.1R$_\text{200m}$.
        Intact satellites begin to contribute near 0.4R$_\text{200m}$.}
    }
    \label{fig:radial-dist}
  \end{figure}

Lastly, we also consider the contributions of each morphological component as a function of distance from the host galaxy.
In Figure \ref{fig:radial-dist} we show the fraction of the accreted stellar mass in radial bins attributed to each morphological type.
The inner regions are mostly made of debris from phase-mixed systems (ignoring \textit{in situ} stellar mass), while stellar streams begin to contribute substantially starting near 0.1$R_\text{200m}$.
Intact satellites, which tend to have recently fallen into the halo (as discussed in further detail in Paper II), begin to contribute near 0.4$R_\text{200m}$.
These characteristic radii are similar to those found in \citet{Genina:2023} for APOSTLE, suggesting that the uptick they find from `surviving dwarfs` near 0.1$R_\text{200m}$ and the increased contribution from `disrupted dwarfs' between 0.2 and 0.6$R_\text{200m}$ (see \citealt{Fattahi:2020} for a similar result in Auriga) are from objects that are in the process of disrupting.
Beyond this radius there is substantial halo-to-halo scatter; while the fraction of accreted material contributed from phase-mixed systems is typically confined to under 15~per~cent, both streams and intact satellites can contribute anywhere from 5 to 90~per~cent of accreted mass (though the median outcome is mostly intact satellites).

\section{Discussion} \label{sec:discussion}

\subsection{Most satellites are streams}

One of the most striking results from our classification analysis is how prevalent streams are in our catalogue of accreted material.
Streams outnumber intact satellites by a factor of two (Figure~\ref{fig:halo-structure}) and most objects with a bound progenitor, which would traditionally be labelled as `satellites', have experienced substantial amounts of disruption (Section \ref{sec:fbound}~and Figure~\ref{fig:fbound-dist}).
Depending on the exact threshold for $f_\text{bound}$, approximately 50 to 70~per~cent of progenitors in Auriga are in the process of actively disrupting.
As a sort of inversion of this result, 80~per~cent of Auriga streams still have a progenitor ($f_\text{bound} > 0$) at the present day (74~per~cent for level 4, 91~per~cent for level 3, see Section \ref{sec:convergence} for discussion on numerical convergence in $f_\text{bound}$ across levels).

Cosmological simulations based on $\Lambda$CDM using both semi-analytic \citep{Li:2010, Barber:2015} and hydrodynamic \citep{Brooks:2014, Wetzel:2016, Sawala:2016, Fattahi:2018} prescriptions consistently invoke tidal disruption to explain properties of the observed Milky Way satellites such as the $M_\ast - V_\text{max}$ relation.
Specifics on the amount of stellar mass lost on an object-by-object basis are rarer.
Figure 12 of \citet{Cooper:2010} shows the distribution of $1 - f_\text{bound}$ for six Aquarius haloes, which bears a striking resemblance to the bottom panel of our Figure \ref{fig:fbound-dist}.
Using the APOSTLE simulations, \citet{Wang:2017} found that $\sim$30~per~cent of satellites lose more than 20~per~cent of their stellar mass after infall, though the central stellar discs in Auriga are more massive than in APOSTLE, affecting the number \citep{Richings:2020} and orbits \citep{Riley:2019} of bound satellites.
\citet{Shipp:2023} found that 61 out of 140 satellites with bound mass $M_\ast \gtrsim 5\times10^5$~M$_\odot$, or 44~per~cent, are progenitors of streams identified in FIRE by \citet{Panithanpaisal:2021}.

These results are striking, considering that of the 13 Milky Way satellites that are above $M_\ast = 5\times10^{5}$~M$_\odot$, \footnote{Based on the \href{https://github.com/apace7/local_volume_database}{Local Volume Database} \citep{Pace:2024}.} only Sagittarius shows clear tidal tails, along with Antlia~II that has a large size and velocity gradient indicative of tidal disruption \citep{Torrealba:2019, Ji:2021}.
In addition, there are only two known Milky Way streams \citep[Sagittarius and OC;][]{Li:2022} originating from satellite galaxies above the same mass scale\footnote{To arrive at this tally, we adopt metallicity measurements from \citet{Li:2022}, Chenab's metallicity for the OC stream, and a cutoff at [Fe/H]~$=-1.8$ \citep{Kirby:2013}. Relaxing any of these criteria could reasonably lead to the inclusion of the Jhelum, Cetus-Palca, and Indus streams \citep[e.g.~for higher mass estimates for Cetus-Palca, see][]{Thomas:2022, Yuan:2022}.}.
However, most of the streams predicted by cosmological simulations are very low surface brightness and would be detected as bound progenitors or evade detection altogether by current imaging surveys, but may be revealed by the upcoming facilities \citep{Wang:2017, Shipp:2023}.
Ultimately, assessing the consistency of the Auriga streams with current observations and making predictions for future surveys will require detailed mock observations in the style of \citet{Shipp:2023} for the Milky Way and \citet{Miro-Carretero:2025} for extragalactic hosts.

\subsection{Streams formed from preprocessing} \label{sec:streams-of-sats}

\begin{figure}
    \includegraphics[width=1.0\linewidth]{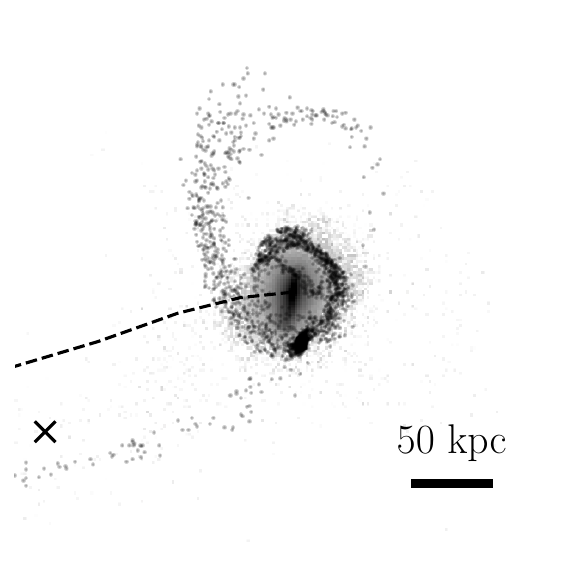}
    \caption{
    Example of a stream that is disrupting around another satellite as it accretes onto the main host.
    The background shade represents the amount of stellar mass from the primary satellite in a pixel, logarithmically scaled.
    Individual points correspond to star particles from the disrupting secondary satellite.
    The dashed line traces the recent trajectory of the primary satellite and the cross indicates the centre of the main host (note that the centre of the primary is 310~kpc from the host and apparent proximity of the tidal debris to the cross is a projection effect).
    \textit{Systems merging with each other prior to falling into the main host, and the streams that naturally form as a result, are included in our sample.}
    }
    \label{fig:stream-of-satellite}
\end{figure}

A surprising outcome from our procedure to identify (Section \ref{sec:particle_lists}) and classify (Section \ref{sec:classifying_streams}) accretion events in Auriga is a population of systems that are not disrupting around the central galaxy, but instead around massive satellites within $R_\text{200m}$ of the Milky Way-mass host.
Group infall onto Milky Way-mass galaxies received new interest when a population of satellites associated with the LMC was identified \citep{Bechtol:2015, Koposov:2015} and subsequently confirmed with 6-D phase space measurements \citep{Kallivayalil:2018, Patel:2020}.
It is now generally accepted that massive satellites accreting onto Milky Way-mass hosts are accompanied by a system of less massive satellites \citep{Li:2008, Wetzel:2015, Jethwa:2016, Santos-Santos:2021}, extending similar analyses for galaxy groups and clusters \citep[e.g.][]{McGee:2009, Wetzel:2013, Joshi:2019} to smaller scales.
\citet{Shao:2018} examined group infall of bound satellites in Auriga, finding that 25 (40)~per~cent of the most massive 11 (50) satellites per halo were accreted in a group in level~4 (3) and that more massive primary satellites bring in larger systems of sub-satellites.

It is straightforward to imagine that many of these systems can lose stellar mass due to the primary's tidal field before falling onto the Milky Way-mass host, forming streams.
This can be viewed as an intermediate step on the process of forming so-called `dwarf stellar haloes' via dwarf-dwarf mergers \citep{Deason:2014, Deason:2022}, just as coherent streams that disrupt around Milky Way-mass hosts eventually phase-mix into the stellar halo.

We show an example of this disruption mechanism from the level~3 run of Au-21 in Figure \ref{fig:stream-of-satellite}.
In this example, a $M_\ast = 3.24 \times 10^{7}$~M$_\odot$ stream is disrupting around a $M_\ast = 1.86 \times 10^{10}$~M$_\odot$ primary that crossed the main host's $R_\text{200c}$ around 3.5~Gyr ago and is currently 310~kpc from the main host, in between $R_\text{200c}$ and $R_\text{200m}$.
This same stream can be seen in a different projection in Figure~\ref{fig:streamportrait}, where in the middle panel for Au-21 there is a stream disrupting around the massive intact satellite in the panel directly above.

These preprocessed systems can appear as outliers in several parameters; objects classified as streams or phase-mixed with recent ($< 2$~Gyr) accretion times onto the main host and large pericentres (see Paper II for more detail).
While we reserve a more in-depth look at stellar streams formed via group infall for future work, it can be useful to have a simple flag to indicate potential preprocessing.
We trace every system backwards in time to see if there are three consecutive snapshots where it is not the most massive halo in its FOF group, ignoring cases where the FOF group is that of the main Milky Way-mass host.
If this condition is satisfied, we label that system as a case of preprocessing\footnote{While we do not examine whether an accretion event actually begins disrupting before accreting onto the main host, this definition does capture all of the instances of preprocessing that we identified visually or as clear outliers in accretion time or dynamical quantities (see Paper II).}.
Across all haloes and resolution levels, 17~per~cent of accretion events satisfy this condition, with an increased prevalence of 28~per~cent for objects classified as streams (18~per~cent for intact, 11~per~cent for phase-mixed).
These fractions agree with those computed for levels 4 and 3 separately, except the value for intact satellites is only 4~per~cent in level~3, possibly reflecting the unique satellite interactions for which those haloes were selected.

Clearly preprocessing of satellites plays a substantial role in producing stellar streams, an effect which has been relatively under-explored.
We emphasise that this mechanism is not rare, with approximately 1 in 4 streams in our sample residing in a more massive halo prior to infall.
It shines light on a tantalising question -- if the Milky Way \citep[$M_\text{200c} \sim 1\times10^{12}$~M$_\odot$, $M_\ast \sim 5\times10^{10}$~M$_\odot$;][]{Bland-Hawthorn:2016} has over 100 stream candidates with 8 streams spectroscopically confirmed to originate from satellite galaxies, where are the streams disrupting around the LMC \citep[$M_\text{200c} \sim 2\times10^{11}$~M$_\odot$, $M_\ast \sim 3\times10^9$~M$_\odot$;][]{Vasiliev:2023}?
Future deep photometry from the Rubin Observatory and Roman Space Telescope could provide exciting answers in the coming decade, potentially adding a new layer to the tangled web of disrupting satellites that comprise the Galactic halo.

\subsection{Shortcomings of labelling accretion events} \label{sec:shortcomings-labelling}

During the process of constructing our visually classified set, it became clear that while we imagine three clearly defined morphological classes, nature (or at least the Auriga haloes) does not seem to care about such desires.
Figure \ref{fig:streamportrait} shows that while most objects are straightforward to classify, others contain overlapping qualities from different labels.
Massive accretion events can lose under 1~per~cent of their stellar mass, but in doing so form nascent tidal tails that contain more stellar mass than the total of some accretion events.
Some phase-mixed objects can have a dominant dense component that is centrally concentrated and spatially smooth, but also contain cold substructure(s) that extends to large radii.
Under the messiness of a cosmological assembly history and a potential that evolves non-adiabatically at early times \citep{Panithanpaisal:2021}, applying a single label to an object may be an oversimplification.

To maintain consistency with other prior studies \citep{Panithanpaisal:2021, Shipp:2023}, we produce a single label based on the classification criteria described in Section \ref{sec:classifying_streams}.
However, even in instances where the science case is motivated by a particular morphological class (e.g.~stream detectability), it may be beneficial to apply an analysis uniformly to all accreted structures, so that edge cases can present themselves in the relevant manner.
Another possible path forward could be to identify whether an individual object has a particular morphological feature\footnote{This should be independent of assessing the detectability of the feature, since the results of such an exercise will inevitably vary based on observational mode.} (tidal tail, shell), ideally still using an automated procedure that is agnostic to the resolution of the simulation.

\subsection{The Auriga haloes in context} \label{sec:auriga-issues}

\begin{figure}
    \includegraphics[width=1.0\linewidth]{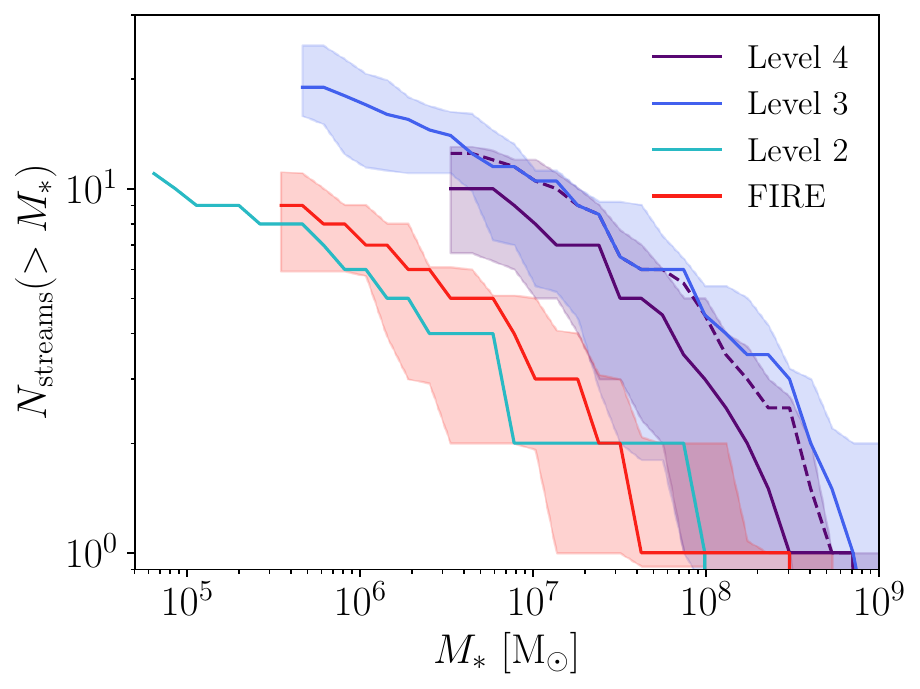}
    \caption{
    Stellar mass functions for stellar streams.
    Solid lines indicate the median and shaded regions the 16th and 84th percentiles, for each Auriga level as well as FIRE \citep{Panithanpaisal:2021}.
    The dashed line indicates the median of the level~4 result for the six haloes that were re-simulated at level~3.
    \textit{While assembly history plays an important role in setting the stream stellar mass function, the Auriga haloes have approximately twice as many streams as FIRE at fixed stellar mass.}
    }
    \label{fig:stellar-mass-function}
\end{figure}

All simulations are imperfect representations of the physical processes that they attempt to capture, and the Auriga simulations are no exception.
Here we discuss some of the caveats that are specific to Auriga and that are worth bearing in mind when digesting our results on disrupting satellites.

The first of these is that the stellar mass-halo mass relation in Auriga is considerably higher than in comparable cosmological simulations \citep[see Figure 2 of][]{Sales:2022} and observational constraints \citep{Behroozi:2013smhmr, Moster:2013}, particularly in the stellar mass range of $10^8 - 10^{10}$~M$_\odot$ (though see \citealt{Posti:2019, Grand:2024} for discussion on the uncertainty in this relation and \citealt{Simpson:2018} for comparisons between the Auriga and Milky Way satellite luminosity functions).
As an example, an Auriga central galaxy with $M_\ast \sim 10^9$~M$_\odot$ can expect to live in a halo of $M_\text{200c} \sim 2 \times 10^{10}$~M$_\odot$, while a galaxy of the same stellar mass in APOSTLE enjoys a much more spacious halo of $M_\text{200c} \sim 10^{11}$~M$_\odot$.
This probably means that the number of objects at this stellar mass scale is slightly higher and such objects experience less dynamical friction in Auriga than in other simulations (since they would occupy more massive haloes, which are rarer in $\Lambda$CDM).
Both of these effects boost the number of massive ($M_\ast > 10^8$~M$_\odot$) streams.

\begin{figure*}
    \includegraphics[width=1.0\linewidth]{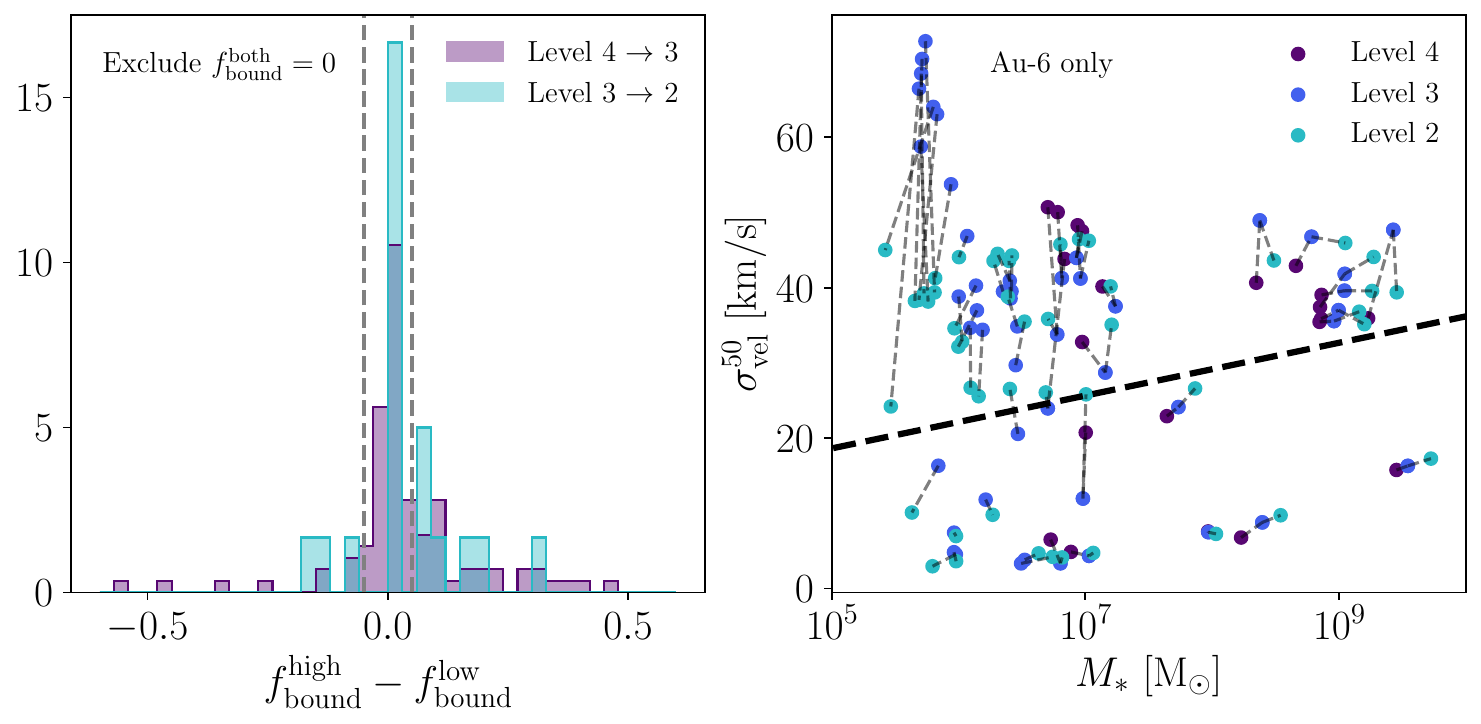}
    \caption{
    Convergence of accreted objects in $f_\text{bound}$ and the stellar mass-velocity dispersion plane used for morphological classification.
    Left: change in $f_\text{bound}$ between objects matched across resolution levels, excluding objects that have no progenitor at both resolutions $(f_\text{bound}^\text{high} = f_\text{bound}^\text{low} = 0)$.
    Dashed lines indicate an absolute change of $\pm5$~per~cent and the y-axis units are arbitrary to normalise both histograms to unity.
    Right: stellar mass-velocity dispersion plane, as presented in Figure~\ref{fig:veldisp-smass}, only showing systems in Au-6 for visual clarity.
    Each point corresponds to a system at a given resolution level, with dashed lines connecting objects to their counterpart one resolution level higher.
    Some objects only begin to be resolved at level~3, while objects that are only resolved in level~2 are not shown.
    \textit{Most accretion events are sufficiently converged within the stellar mass-velocity dispersion plane, lending confidence to our stream/phase-mixed classification, while we find a modest preferential increase in \textnormal{$f_\text{bound}$} with increasing resolution.}
    }
    \label{fig:convergence}
\end{figure*}

Even given a fixed stellar mass-halo mass relation, it is clear from our results (e.g.~Figures \ref{fig:streamportrait} and \ref{fig:halo-structure}) that the number of streams -- and their overall contribution to the accreted material -- is sensitive to the recent assembly history.
The sample of haloes simulated at level~4 likely captures a representative range of histories for isolated Milky Way-mass haloes (indeed, this is the intention of the halo selection process described in Section \ref{sec:sims}).
However, this does not need to be true for level~3, whose haloes were often selected for re-simulation due to their massive central stellar discs (Au-16, Au-24) or interesting interactions with massive satellites (Au-21, Au-23, Au-27), features that could prefer `noisier' assembly histories with more streams.
Prospects for a representative sample of Auriga haloes with high stellar mass resolution should be improved by the ongoing Auriga Superstars project \citep[Pakmor et al.~in preparation, Fragkoudi et al.~in preparation]{Grand:2023}.
While we report results for the full range of haloes available, making detailed comparisons with observations or across different simulations must be interpreted with an eye on the effect of accretion history.

As an example, in Figure \ref{fig:stellar-mass-function} we compare the stellar mass functions of streams in our sample at different resolution levels, as well as to those identified in FIRE \citep{Panithanpaisal:2021} which are similar in mass resolution to our level~3.
The apparent disagreement between level~4 and 3 arises from the assembly histories of the level~3 haloes; the median of the level~4 runs for only Au-(6, 16, 21, 23, 24, 27), shown as a dashed line, agrees with the median for level~3.
The apparent disagreement between level~2 and the other Auriga levels is due to the quiet merger history of Au-6; the lower resolution runs of the same halo have similarly low numbers of streams relative to the rest of the Auriga sample (see Appendix \ref{app:au6-streams}).
Finally, at fixed stellar mass we find that there are approximately twice as many streams in Auriga as in FIRE, an effect which is not caused by differences in our definitions of a stream (see Appendix \ref{app:fire-classification}).
It is possible that a combination of different assembly histories and stellar mass-halo mass relations could alleviate the difference in stream mass functions between FIRE and Auriga, or it could still point towards interesting differences in galaxy formation physics (see Paper II for further discussion on this point).
Stream mass functions in comparable suites of cosmological simulations of Milky Way-mass haloes might be illuminating on this point.

Finally, it is possible that numerical issues related to the finite mass resolution of cosmological simulations could affect how satellites disrupt.
Recent studies using tailored $N$-body simulations have highlighted that subhaloes can artificially disrupt due to an insufficient number of particles or force resolution \citep{vandenBosch:2018, Errani:2020, Errani:2021}, though such experiments do not capture the important effects of preprocessing \citep{He:2025} or the tidal field of a central massive disk \citep{Garrison-Kimmel:2017, Kelley:2019, Wang:2025}.
In addition, spurious heating of galaxies due to unequal dark matter and star particle masses or insufficient total number of particles \citep{Ludlow:2019, Ludlow:2020, Ludlow:2021, Ludlow:2023} could make satellites easier to disrupt into streams; the impact of spurious heating can be tested by repeating this analysis with forthcoming simulations that explicitly account for this effect (e.g.~\textsc{colibre}; Schaye et al.~in preparation).

\subsection{Convergence across resolution} \label{sec:convergence}

Having a sample of accretion events simulated at different resolution levels characterised in a uniform manner provides us an opportunity to study numerical convergence for accreted structures in Auriga.
We focus on systematic trends for individual objects as resolution increases, rather than assessing if every system is reproduced perfectly at each resolution (see Section~\ref{sec:matching}) or global properties of the accreted material (see Table~\ref{tab:haloes} and Section~\ref{sec:halostructure}).

We find that there is a slight systematic trend that satellites are better at retaining stars with increasing resolution.
In Figure~\ref{fig:convergence} we show the change in $f_\text{bound}$ for objects that have a progenitor in either resolution level.
For the 65 systems matched from level~4 to 3 with an absolute change in $f_\text{bound}$ of at least 1~per~cent\footnote{We note that this particular statistic emphasises changes in $f_\text{bound}$ with resolution. Many objects consistently have $f_\text{bound} = 0$ or $=1$ (55~per~cent from level~4 to 3 and 61~per~cent from level~3 to 2).}, this difference is $f_\text{bound}^\text{high} - f_\text{bound}^\text{low} = 0.04^{+0.15}_{-0.08}$.
From level~3 to 2, the 11 matched systems satisfying the same condition have $f_\text{bound}^\text{high} - f_\text{bound}^\text{low} = 0.07^{+0.10}_{-0.17}$.
In addition, 13 (7) accretions from level~4 to 3 (3 to 2) have no bound progenitor at lower resolution but some progenitor ($f_\text{bound} > 0$) at higher resolution.
Finally, we can assess the impact of resolution on actively disrupting satellites by considering objects that are classified as streams in both level~4 and level~3.
Of these 60 streams, 50 (83~per~cent) have $f_\text{bound} > 0$ in level 4 and 59 (98 per cent) in level 3.
The nine streams that gain a progenitor in level 3 have $f_\text{bound}$ between 0.0006 and 0.08, suggesting that they are in a similar dynamical state in level 4 but the progenitor is not recovered at lower resolution.
While substantial changes (over $\sim$30~per~cent) in $f_\text{bound}$ are often instances of different accretion timing and number of experienced pericentric passages (see Section \ref{sec:matching}), these results suggest some impact of numerical resolution on individual systems \citep{vandenBosch:2018, Errani:2020}.
However, we note that the bulk of systems appear remarkably similar (Figure~\ref{fig:matching-stellar-maps}) and that global properties of all accreted stellar mass are largely preserved (Table~\ref{tab:haloes} and Figure~\ref{fig:halo-structure}).

In Figure~\ref{fig:convergence} we also show the distribution of accretions in the same $M_\ast - \sigma_\text{vel}^{50}$ plane that we use to classify systems as streams or phase-mixed, but now connecting objects matched in Au-6 at all resolution levels (similar trends apply to other haloes simulated at level~3).
We only include objects that have a lower resolution counterpart, such that some systems are only represented in levels 3 and 2.
Structures that are only resolved in level~2 are omitted.
While most accretions (especially at higher mass) do not change substantially in this view, two systematic trends with resolution emerge.
The first is that individual systems, particularly those with $M_\ast > 10^7$~M$_\odot$, have higher stellar masses with increasing resolution, an effect that has been noted in previous Auriga studies \citep{Grand:2017, Grand:2021}.
The second is that phase-mixed systems at the lower mass end of a given resolution level ($\lesssim 10^7$~M$_\odot$ for level~4, $\lesssim 10^6$~M$_\odot$ for level~3) tend to have substantially lower $\sigma_\text{vel}^{50}$ in their higher resolution counterparts.
This is an \textit{apparent} difference (not attributable to the Auriga model) due to our requirement that at least $k=7$ neighbours are used when calculating the local velocity dispersion at lower resolution; the change comes from switching to 1~per~cent of star particles in the object at higher resolutions.
We note that even with both of these effects, the variations for individual systems in this plane do not typically result in crossing the stream/phase-mixed separation line, lending confidence to our classification method.

\section{Summary} \label{sec:summary}

We have used the Auriga suite of cosmological simulations to identify and characterise accreted satellites at all stages of tidal disruption around Milky Way-mass haloes.
We establish a uniform set of criteria for automatically classifying accretion events as intact satellites, stellar streams, or phase-mixed systems based on the fraction of stellar mass bound to the progenitor at the present day and the median local velocity dispersion of the structure.
This classification method is calibrated to reproduce results for a visually classified sample and comfortably scales to different simulation resolution levels.
We also match accretion events in the same halo across different resolution levels, leading to new insights into the effect of numerical resolution for disrupting satellites.
In total, we identify 175 intact satellites, 417 stellar streams, and 802 phase-mixed systems across all haloes and resolution levels.

Our key findings are the following:
\begin{enumerate}
    \item Satellites that are actively unravelling due to tidal forces are the norm, not the exception. Of all systems that have a bound progenitor, 67~per~cent have $f_\text{bound} < 0.97$ -- our threshold for a satellite to no longer be considered intact -- while 53~per~cent satisfy a more stringent $f_\text{bound} < 0.8$ (Figure~\ref{fig:fbound-dist}). Approximately 80~per~cent of streams have a bound progenitor at the present day.
    \item The stellar morphologies of individual objects are remarkably converged across resolution, often with only minor differences in the orbital phase at the present day (Figure~\ref{fig:matching-stellar-maps}). Instances where this is not the case have matching dark matter distributions on large scales but mismatched stellar distributions due to slightly different accretion times and number of experienced pericentric passages (Figure~\ref{fig:dm_mismatches}).
    \item When viewing the structure of accreted material under these three stages of disruption, the number of objects systematically increases with disruption stage but most of the stellar mass is contributed by intact or phase-mixed systems, with substantial variety across haloes in the fraction of stellar mass in each stage (Figures~\ref{fig:barcharts} and \ref{fig:halo-structure}). Streams are substantial contributors to the stellar halo at intermediate distances from the host centre, between $\sim$0.1 and $\sim$0.7$R_\text{200m}$ (Figure~\ref{fig:radial-dist}) corresponding to $\sim$35 and $\sim$250~kpc for the Milky Way \citep[adopting the $R_\text{200m}$ estimate from][]{Deason:2020}.
    \item Massive accretion events in Auriga can have their own systems of disrupting satellites formed from preprocessing prior to accreting onto the main host (Figure~\ref{fig:stream-of-satellite}). Streams in Auriga are more likely to experience preprocessing than intact or phase-mixed systems, suggesting that this mechanism plays an important role in setting disruption rates around Milky Way-mass haloes.
    \item Auriga predicts twice as many stellar streams as FIRE at fixed stellar mass (Figure~\ref{fig:stellar-mass-function}), though it is difficult to disentangle effects of halo assembly history from the difference in galaxy formation physics between the two suites.
\end{enumerate}

Our uniform characterisation of every accretion event in Auriga provides an unprecedented dataset for studying disrupting satellites in a cosmological context.
We note that we have not yet analysed the detectability of the Auriga streams, many of which are likely too faint for  current imaging \citep[e.g.][]{Shipp:2023} but could be revealed with future facilities including LSST, Euclid, and ARRAKIHS.
To facilitate comparisons with both observations and other simulations, we make this catalogue of accretion events publicly available (see the Data Availability Statement and Appendix \ref{app:datatables}).
In addition to the orbital properties examined in Paper II, there are ongoing efforts to study the observational detectability, star formation histories, and chemical properties of this sample across the different stages of disruption.

\section*{Acknowledgements}
We thank Emily Cunningham, Denis Erkal, Danny Horta, Nondh Panithanpaisal, Robyn Sanderson, Eugene Vasiliev, and Anna Wright for enlightening discussions, as well as an anonymous referee for helpful suggestions regarding its presentation.
AHR thanks Taylor Swift's \textit{The Tortured Poets Department} and Lindsey Stirling's \textit{Duality} for providing a soundscape condusive to writing this manuscript.
This project was developed in part at the Streams24 meeting hosted at Durham University.

AHR is supported by a Research Fellowship from the Royal Commission for the Exhibition of 1851 and by STFC through grant ST/T000244/1.
NS was supported by an NSF Astronomy and Astrophysics Postdoctoral Fellowship under award AST-2303841.
RB is supported by the UZH Postdoc Grant, grant no.~FK-23116 and the SNSF through the Ambizione Grant PZ00P2\_223532.
AF and STB are supported by a UKRI Future Leaders Fellowship (grant no MR/T042362/1).
KAO acknowledges support by the Royal Society through Dorothy Hodgkin Fellowship DHF/R1/231105 and by STFC through grant ST/T000244/1.
FF is supported by a UKRI Future Leaders Fellowship (grant no. MR/X033740/1).
FAG acknowledges funding from the Max Planck Society through a “PartnerGroup” grant.
FAG acknowledges support from ANID FONDECYT Regular 1211370, the ANID Basal Project FB210003 and the HORIZON-MSCA-2021-SE-01 Research and innovation programme under the Marie Sklodowska-Curie grant agreement number 101086388.
RJJG is supported by an STFC Ernest Rutherford Fellowship (ST/W003643/1).
FM acknowledges funding from the European Union - NextGenerationEU under the HPC project `National Centre for HPC, Big Data and Quantum Computing' (PNRR - M4C2 - I1.4 - CN00000013 – CUP J33C22001170001).
This work was supported by collaborative visits funded by the Cosmology and Astroparticle Student and Postdoc Exchange Network (CASPEN).

This work used the DiRAC@Durham facility managed by the Institute for Computational Cosmology on behalf of the STFC DiRAC HPC Facility (www.dirac.ac.uk).
The equipment was funded by BEIS capital funding via STFC capital grants ST/K00042X/1, ST/P002293/1, ST/R002371/1 and ST/S002502/1, Durham University and STFC operations grant ST/R000832/1.
DiRAC is part of the National e-Infrastructure.
This research used resources of the Argonne Leadership Computing Facility, a U.S. Department of Energy (DOE) Office of Science user facility at Argonne National Laboratory and is based on research supported by the U.S. DOE Office of Science-Advanced Scientific Computing Research Program, under Contract No. DE-AC02-06CH11357.
This work used the Freya computer cluster at the Max Planck Institute for Astrophysics.

For the purpose of open access, the author has applied a Creative Commons Attribution (CC BY) licence to any Author Accepted Manuscript version arising from this submission.

This research made extensive use of \href{https://arxiv.org/}{arXiv.org} and NASA's Astrophysics Data System for bibliographic information.

\section*{Software}
This research made use of the Python programming language, along with many community-developed or maintained software packages including:
\begin{itemize}
    \item Astropy \citep{Astropy:2013, Astropy:2018, Astropy:2022}
    \item CMasher \citep{cmasher, cmasher1.8.0}
    \item Cython \citep{cython}
    \item h5py \citep{h5py, h5py3.7.0}
    \item Jupyter \citep{ipython, jupyter}
    \item Matplotlib \citep{matplotlib}
    \item NumPy \citep{numpy}
    \item Pandas \citep{pandas, pandas1.5.0}
    \item Parsl \citep{parsl}
    \item Scikit-learn \citep{scikit-learn, scikit-learn-api, scikit-learn1.1.2}
    \item SciPy \citep{scipy, scipy1.9.1}
\end{itemize}
We also thank the maintainers of the arepo-snap-util package.
Parts of the results in this work make use of the colormaps in the CMasher package.
Software citation information aggregated using \href{https://www.tomwagg.com/software-citation-station/}{The Software Citation Station} \citep{software-citation-station-paper, software-citation-station-zenodo}.

\section*{Data Availability}

Halo catalogs, merger trees, and particle data (Sections \ref{sec:sims} and \ref{sec:particle_lists}) for Auriga levels 3 and 4 are publicly available \citep[detailed in the Auriga project data release;][]{Grand:2024} to download via the Globus platform,\footnote{\href{https://wwwmpa.mpa-garching.mpg.de/auriga/data.html}{https://wwwmpa.mpa-garching.mpg.de/auriga/data}} while the analogous data products for Auriga level~2 will be shared upon reasonable request.
The catalogue of accreted structures characterised in this article (Sections \ref{sec:classifying_streams} and \ref{sec:matching}) is available in Appendix~\ref{app:datatables} and on \href{https://wwwmpa.mpa-garching.mpg.de/auriga/dataspecs.html#Highleveldata}{the Auriga webpage}.


\bibliographystyle{mnras}
\bibliography{main, software}


\appendix

\section{Auriga systems with the Panithanpaisal et al. (2021) method} \label{app:fire-classification}

\begin{figure}
    \includegraphics[width=1.0\linewidth]{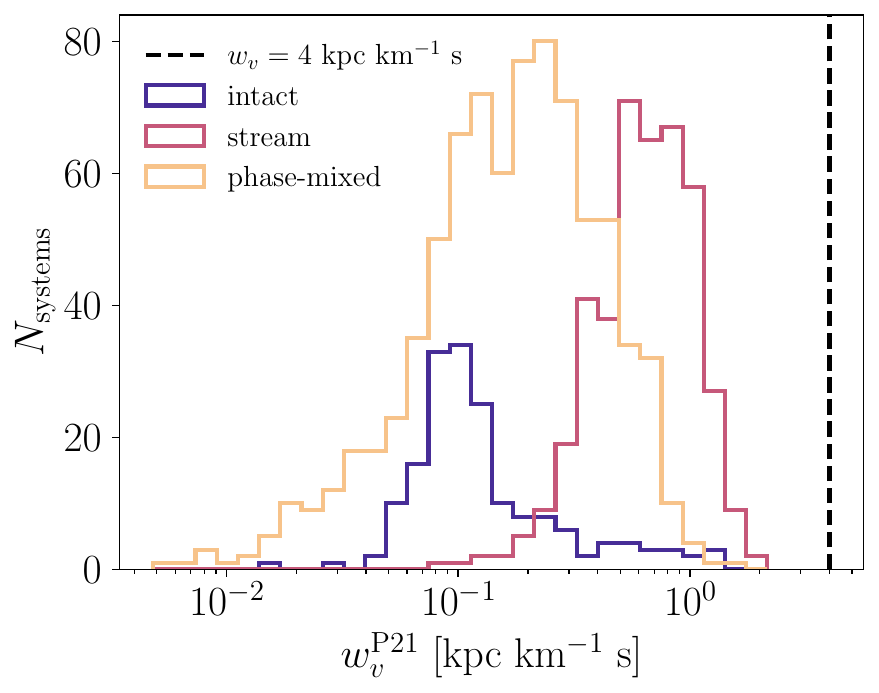}
    \caption{
    Distribution of $w_v^\text{P21}$ for Auriga accretions from our visually classified sample, separated by morphology based on our automated classification scheme (Section \ref{sec:classifying_streams}).
    This quantity is defined as the ratio of position and velocity dispersion $\sigma_\text{pos} / \sigma_\text{vel}$ across all star particles in a given object, as used in \citet{Panithanpaisal:2021}.
    We also indicate $w_v = 4$~kpc~km$^{-1}$~s that we apply uniformly to all objects in this work.
    \textit{The relative weight of position and velocity information in their method varies by over two orders of magnitude, in a way that maps non-trivially onto how disrupted the system is.}
    }
    \label{fig:wv-nat-histogram}
\end{figure}

Even though our approach to classifying morphologies shares many characteristics with that used in \citet{Panithanpaisal:2021}, there are notable differences that we describe in Section~\ref{sec:nondh-comparison}.
Here we examine whether these differences result in different classifications on the same data, which would make comparisons between our samples substantially more challenging.

First, we briefly motivate our choice to apply a uniform velocity-to-distance scaling $w_v$ when finding nearest neighbours for the local velocity dispersion calculation.
\citet{Panithanpaisal:2021} choose this weighting based on the ratio of the \textit{global} dispersions in position and velocity separately, resulting in a $w_v$ that differs from object to object.
In Figure~\ref{fig:wv-nat-histogram}, we show the distribution of this value for the Auriga accretions from our visually classified sample (across all resolutions).
It is clear that this method of choosing $w_v$ means that the metric for defining nearest neighbours, a step in the automated classification procedure, depends on accretion morphology in a non-trivial manner.
While this may be a justifiable choice, we prefer to adopt a single value of $w_v$ applied to all of the data.

This choice, as well using $f_\text{bound}$ rather than the spatial extent of the object to define intact satellites, could result in substantially different classifications of the same system.
To test this, we apply their spatial separation criteria and method of varying $w_v$ to the level~4 systems in our visually classified sample, including re-calibrating our SVM on the resulting local velocity dispersions.
We restrict to the visually classified sample to also compare to our by-eye classifications, and to level~4 haloes since it is faster to explicitly calculate the maximum particle separation with fewer star particles.
We also exclude systems with $M_\ast > 10^9$~M$_\odot$, as this is outside the range that \citet{Panithanpaisal:2021} consider.
Since the level~4 runs are lower resolution than the FIRE haloes (which are comparable in particle mass to level~3), we still use our technique of anchoring the number of nearest neighbours $k$ to 1~per~cent of particles in the system, which largely accounts for the resolution difference.

\begin{figure}
    \includegraphics[width=1.0\linewidth]{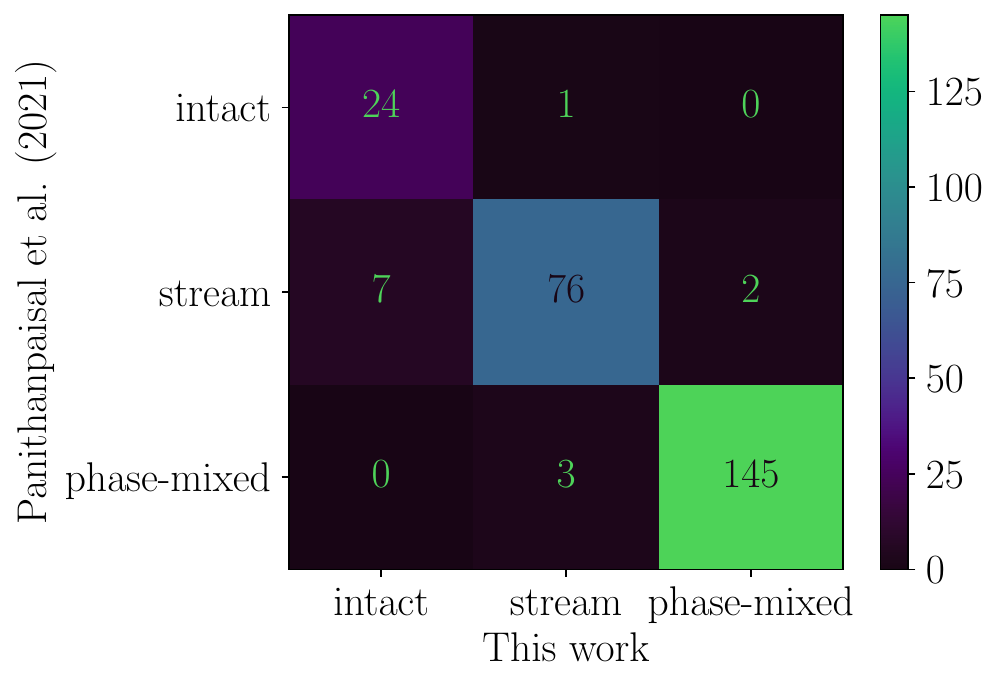}
    \caption{
    Comparison of the automated classification method presented in this work and that in \citet{Panithanpaisal:2021}.
    Both methods were applied to the sample of level~4 Auriga accretions from our visually classified sample.
    \textit{Substantial differences in stream populations between FIRE and Auriga are likely not driven by classification methods or definitions.}
    }
    \label{fig:fire-auriga-confusion-matrix}
\end{figure}

Figure~\ref{fig:fire-auriga-confusion-matrix} shows a confusion matrix comparing the result of this classification exercise to our method detailed in Section~\ref{sec:classifying_streams}.
Overall, the two methods agree quite well, with 95~per~cent of systems having the same result.
The instances of disagreement are often edge cases in one classification scheme or the other.
Seven objects have a maximum pairwise separation over 120~kpc but have only lost 1-2~per~cent of their stellar mass (sometimes with only a few very distant particles), above our $f_\text{bound}$ threshold.
One system has no bound progenitor and clear tidal tails, but the maximum pairwise separation is 118~kpc, below the 120~kpc limit set in \citet{Panithanpaisal:2021}.
Objects that are classified as streams in one method and phase-mixed in the other often have a smaller number of particles ($\sim$150 to $\sim$600) and/or lie near the edge of the stream/phase-mixed classification plane of both methods.

Ultimately, we conclude that substantial differences in the disrupting populations between this study and \citet{Panithanpaisal:2021}, our Figure~\ref{fig:stellar-mass-function} for example, are too large to be explained by the different techniques of classifying systems.

\section{Catalogue of accreted structures} \label{app:datatables}

In Table~\ref{tab:accretion-props} we present the complete catalogue of accretion events identified in Section~\ref{sec:particle_lists}, as well as information related to morphological classification (Section~\ref{sec:classifying_streams}), matching to higher resolution (Section~\ref{sec:matching}), and preprocessing (Section~\ref{sec:streams-of-sats}).
The system IDs match the `accreted particle lists' in the Auriga public data release \citep{Grand:2024}.

\begin{table*}
\centering
\input{tables/accretion-props}
    \caption{
    Catalogue of accretion events and their properties analysed in this work, sorted by level, then halo number, then stellar mass.
    For brevity, we only show systems for the level~3 run of Au-6 in this manuscript.
    We provide the halo number, resolution level, and system ID that in combination uniquely identify an object (IDs alone are not guaranteed to be unique); morphological classification (Section \ref{sec:classifying_streams}); total stellar mass ($M_\ast$), including bound progenitor if still present at $z=0$; number of star particles ($N_\text{part}$); fraction of stellar mass bound to the progenitor ($f_\text{bound}$); whether the object experienced preprocessing (Section \ref{sec:streams-of-sats}); and the matched ID of the same object at one resolution level higher, as well as a quality assessment of that match.
    A machine-readable table with the full catalogue is available as supplementary material.
    }
    \label{tab:accretion-props}
\end{table*}

\section{Stream stellar mass functions for Au-6}
\label{app:au6-streams}

\begin{figure}
    \includegraphics[width=1.0\linewidth]{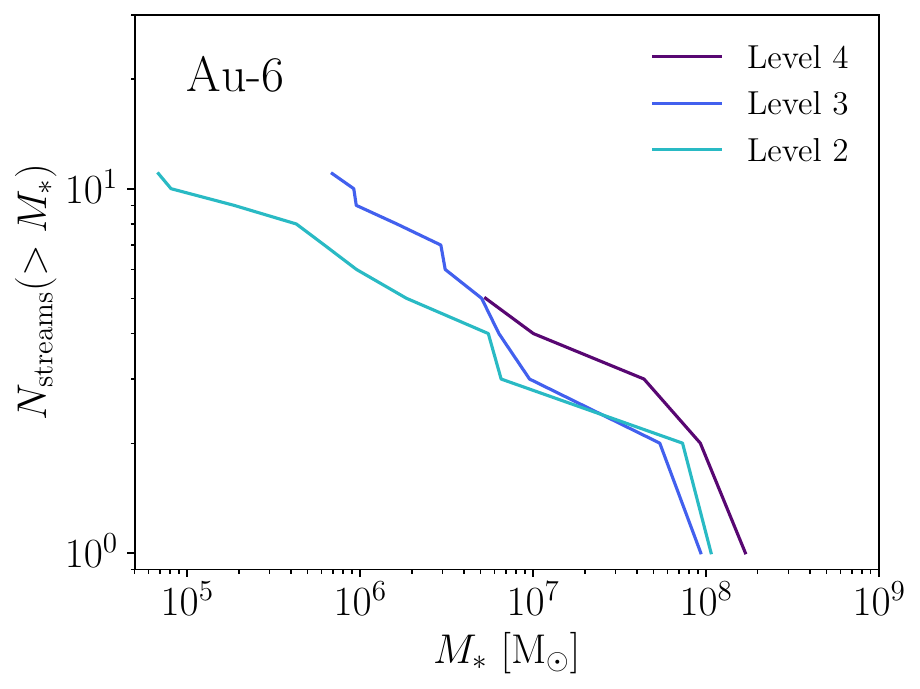}
    \caption{
    Stellar mass functions for stellar streams, focusing only on Au-6 at different resolution levels (see Figure~\ref{fig:stellar-mass-function} for all Auriga haloes).
    The apparent disagreement between level~4 and the other levels at $M_\ast > 10^7$~M$_\odot$ and between levels~3~and~2 at $M_\ast < 3\times 10^6$~M$_\odot$ is due to matched objects having different classifications at different levels (see text of Appendix~\ref{app:au6-streams} for details).
    }
    \label{fig:stellar-mass-function-au6}
\end{figure}

In Figure~\ref{fig:stellar-mass-function-au6} we present the stellar mass function for streams for Au-6 at various resolution levels, as a complement to Figure~\ref{fig:stellar-mass-function}.
At first glance, the apparent disagreement between level~4 and the other levels at $M_\ast > 10^7$~M$_\odot$ and between levels~3~and~2 at $M_\ast < 3\times 10^6$~M$_\odot$ suggest that these stream mass functions are not converged across resolution.
However, this is typically due to individual accretion events having intact or phase-mixed classifications at other levels.
In particular:
\begin{itemize}
    \item The most massive stream in level~4 is classified as intact in levels~3~and~2. This is the same system shown in the leftmost panel of Figure~\ref{fig:matching-stellar-maps}.
    \item Three streams in level~3 are classified as phase-mixed in level~2. All three systems can be seen crossing the stream/phase-mixed boundary in Figure~\ref{fig:convergence}, while the most massive is the case highlighted in Figure~\ref{fig:matching-stellar-maps} (third panel from the right) and Figure~\ref{fig:dm_mismatches} (bottom panels). All level~2 streams with $M_\ast > 4\times10^5$~M$_\odot$ are also streams in level~3, while objects below this mass do not appear in the level~3 catalogues.
\end{itemize}
These cases arise from looking at a single halo with a strict definition for `stream', which will inevitably lead to boundary cases.
Since Au-6 has a very quiet merger history with few mergers overall, variations are relatively impactful on the stream stellar mass function.
These effects are mitigated when considering several haloes at once, as is the case in Figure~\ref{fig:stellar-mass-function}.


\bsp	
\label{lastpage}
\end{document}

%% file: commands.tex
\usepackage{soul}
\usepackage{amsmath}
\usepackage{xspace}
\usepackage{xifthen}
\usepackage{eso-pic}

\definecolor{forestgreen}{HTML}{228B22}
\definecolor{urlblue}{HTML}{000000}

\mathchardef\mhyphen="2D

\newlength{\dhatheight}

\newcommand{\bandvar}[2][]{%
  \ifthenelse{\isempty{#1}}{\var{#2}}{\var{#2\_#1}}%
}

\newcommand{\var}[1]{\ensuremath{\texttt{\MakeUppercase{#1}}}\xspace}

\providecommand\physrep{\ref@jnl{Phys.~Rep.}}%
\providecommand\apjs{\ref@jnl{ApJS}}%
\providecommand{\jcap}{\ref@jnl{JCAP}}%

%% file: tables/halo-props.tex
\begin{tabular}{ccccccccccccc}
\hline
Halo & $M_\text{200c}$ & $R_\text{200c}$ & $R_\text{200m}$ & $M_\ast^\text{in situ}$ & $M_\ast^\text{acc}$ & $R_{50}^\text{in situ}$ & $N_\text{intact}$ & $N_\text{stream}$ & $N_\text{phase-mixed}$ & $f_{\text{acc},\ast}^\text{intact}$ & $f_{\text{acc},\ast}^\text{stream}$ & $f_{\text{acc},\ast}^\text{phase-mixed}$ \\
 & ($10^{12}$~M$_\odot$) & (kpc) & (kpc) & ($10^{10}$~M$_\odot$) & ($10^{10}$~M$_\odot$) & (kpc) \\
\hline
\multicolumn{13}{c}{All resolution levels} \\
\hline
Au-6-L4 & 1.044 & 214 & 340 & 4.796 & 0.926 & 7.58 & 2 & 5 & 17 & 0.310 & 0.035 & 0.655 \\
Au-6-L3 & 1.015 & 212 & 338 & 5.801 & 1.175 & 4.70 & 5 & 11 & 34 & 0.321 & 0.015 & 0.664 \\
Au-6-L2 & 1.024 & 212 & 340 & 6.495 & 1.726 & 6.94 & 7 & 11 & 57 & 0.335 & 0.011 & 0.654 \\
\hline
\multicolumn{13}{c}{Levels 4 and 3} \\
\hline
Au-16-L4 & 1.503 & 241 & 381 & 6.290 & 1.695 & 12.58 & 4 & 12 & 17 & 0.526 & 0.082 & 0.392 \\
Au-16-L3 & 1.504 & 242 & 381 & 8.393 & 2.142 & 9.44 & 8 & 19 & 45 & 0.510 & 0.080 & 0.410 \\
Au-21-L4 & 1.451 & 239 & 405 & 6.995 & 3.557 & 7.37 & 8 & 13 & 15 & 0.386 & 0.281 & 0.333 \\
Au-21-L3 & 1.415 & 237 & 403 & 7.289 & 4.716 & 7.62 & 10 & 24 & 25 & 0.409 & 0.283 & 0.308 \\
Au-23-L4 & 1.575 & 245 & 387 & 8.642 & 1.238 & 5.19 & 6 & 13 & 24 & 0.022 & 0.154 & 0.825 \\
Au-23-L3 & 1.504 & 242 & 382 & 7.896 & 1.555 & 7.62 & 8 & 19 & 52 & 0.017 & 0.148 & 0.834 \\
Au-24-L4 & 1.492 & 241 & 391 & 6.709 & 1.477 & 6.14 & 7 & 12 & 23 & 0.346 & 0.251 & 0.403 \\
Au-24-L3 & 1.468 & 240 & 389 & 7.733 & 1.974 & 8.25 & 8 & 17 & 52 & 0.356 & 0.011 & 0.633 \\
Au-27-L4 & 1.745 & 254 & 411 & 8.995 & 1.843 & 6.21 & 4 & 13 & 22 & 0.240 & 0.098 & 0.662 \\
Au-27-L3 & 1.696 & 251 & 406 & 8.715 & 2.298 & 6.31 & 5 & 27 & 50 & 0.253 & 0.179 & 0.568 \\
\hline
\multicolumn{13}{c}{Level 4 only} \\
\hline
Au-2-L4 & 1.915 & 262 & 425 & 8.091 & 3.072 & 14.04 & 8 & 11 & 28 & 0.559 & 0.026 & 0.415 \\
Au-3-L4 & 1.458 & 239 & 380 & 7.477 & 1.564 & 9.85 & 4 & 8 & 12 & 0.199 & 0.036 & 0.766 \\
Au-4-L4 & 1.409 & 236 & 393 & 5.506 & 3.388 & 7.13 & 3 & 16 & 17 & 0.006 & 0.066 & 0.928 \\
Au-5-L4 & 1.186 & 223 & 354 & 6.356 & 0.876 & 3.95 & 4 & 13 & 16 & 0.108 & 0.169 & 0.723 \\
Au-7-L4 & 1.120 & 219 & 358 & 3.859 & 2.079 & 5.05 & 2 & 10 & 14 & 0.012 & 0.061 & 0.927 \\
Au-8-L4 & 1.081 & 216 & 345 & 3.302 & 2.252 & 11.93 & 5 & 5 & 13 & 0.713 & 0.191 & 0.096 \\
Au-9-L4 & 1.050 & 214 & 345 & 5.875 & 0.429 & 3.15 & 0 & 10 & 21 & 0.000 & 0.054 & 0.946 \\
Au-10-L4 & 1.047 & 214 & 338 & 5.916 & 0.512 & 1.90 & 6 & 6 & 15 & 0.239 & 0.245 & 0.516 \\
Au-12-L4 & 1.093 & 217 & 346 & 5.377 & 1.484 & 4.56 & 2 & 10 & 14 & 0.033 & 0.334 & 0.633 \\
Au-13-L4 & 1.189 & 223 & 369 & 5.830 & 1.470 & 2.39 & 5 & 12 & 14 & 0.377 & 0.288 & 0.335 \\
Au-14-L4 & 1.657 & 249 & 392 & 9.628 & 2.251 & 4.58 & 2 & 10 & 16 & 0.053 & 0.074 & 0.873 \\
Au-15-L4 & 1.222 & 225 & 360 & 3.697 & 2.194 & 6.94 & 6 & 10 & 13 & 0.626 & 0.118 & 0.256 \\
Au-17-L4 & 1.028 & 213 & 336 & 7.725 & 0.319 & 1.91 & 4 & 10 & 21 & 0.120 & 0.477 & 0.403 \\
Au-18-L4 & 1.221 & 225 & 435 & 8.003 & 6.091 & 4.05 & 12 & 17 & 24 & 0.929 & 0.018 & 0.053 \\
Au-19-L4 & 1.209 & 225 & 359 & 4.786 & 1.904 & 7.10 & 4 & 8 & 15 & 0.255 & 0.118 & 0.627 \\
Au-20-L4 & 1.249 & 227 & 385 & 3.658 & 3.458 & 6.10 & 5 & 13 & 16 & 0.363 & 0.144 & 0.494 \\
Au-22-L4 & 0.926 & 205 & 331 & 5.999 & 0.363 & 1.97 & 3 & 5 & 22 & 0.421 & 0.125 & 0.453 \\
Au-25-L4 & 1.221 & 225 & 353 & 3.433 & 2.035 & 9.02 & 3 & 9 & 11 & 0.017 & 0.920 & 0.063 \\
Au-26-L4 & 1.564 & 245 & 406 & 9.880 & 2.111 & 2.17 & 7 & 14 & 21 & 0.285 & 0.015 & 0.699 \\
Au-28-L4 & 1.605 & 247 & 390 & 8.465 & 2.630 & 3.27 & 3 & 12 & 13 & 0.015 & 0.051 & 0.934 \\
Au-29-L4 & 1.542 & 244 & 410 & 6.518 & 5.287 & 6.38 & 5 & 2 & 23 & 0.267 & 0.006 & 0.727 \\
Au-30-L4 & 1.108 & 218 & 354 & 3.549 & 3.412 & 3.57 & 0 & 10 & 10 & 0.000 & 0.690 & 0.310 \\
\hline
\end{tabular}

%% file: tables/accretion-props.tex
\begin{tabular}{cccccccccc}
\hline
Halo & Level & ID & Morphology & $\log_{10}(M_\ast / M_\odot)$ & $N_\text{part}$ & $f_\text{bound}$ & Preprocessed & Match ID & Match quality \\
\hline
\multicolumn{10}{c}{$\cdots$} \\
6 & 3 & 176 & intact & 9.54 & 808418 & 0.997 & False & 232 & MATCHED \\
6 & 3 & 22562 & phase-mixed & 9.43 & 641336 & 0.000 & False & 433688 & MATCHED \\
6 & 3 & 144 & phase-mixed & 9.05 & 267459 & 0.000 & False & 102240 & MATCHED \\
6 & 3 & 2518 & phase-mixed & 9.04 & 264522 & 0.000 & False & 91904 & MATCHED \\
6 & 3 & 2333 & phase-mixed & 9.00 & 237550 & 0.000 & False & 3612 & MATCHED \\
6 & 3 & 151 & phase-mixed & 8.96 & 223356 & 0.000 & False & 184 & MATCHED \\
6 & 3 & 92 & phase-mixed & 8.78 & 143273 & 0.000 & False & 91897 & MATCHED \\
6 & 3 & 175 & intact & 8.40 & 56750 & 0.990 & False & 231 & MATCHED \\
6 & 3 & 40792 & phase-mixed & 8.38 & 54599 & 0.000 & False & 660895 & MATCHED \\
6 & 3 & 2763 & stream & 7.97 & 21637 & 0.410 & False & 3372 & MATCHED \\
6 & 3 & 2729 & stream & 7.73 & 12637 & 0.001 & False & 102455 & MATCHED \\
6 & 3 & 13801 & phase-mixed & 7.42 & 6394 & 0.000 & False & -- & UNMATCHED \\
6 & 3 & 40886 & phase-mixed & 7.24 & 3944 & 0.000 & False & 484645 & MATCHED \\
6 & 3 & 33060 & phase-mixed & 7.16 & 3332 & 0.000 & True & 403733 & MATCHED \\
6 & 3 & 365 & intact & 7.03 & 2361 & 0.998 & False & 3105 & MATCHED \\
6 & 3 & 13857 & stream & 6.98 & 2161 & 0.134 & False & 392843 & MATCHED \\
6 & 3 & 82 & phase-mixed & 6.96 & 2067 & 0.000 & False & 151 & MATCHED \\
6 & 3 & 450129 & phase-mixed & 6.93 & 1879 & 0.000 & False & 1346443 & MATCHED \\
6 & 3 & 441235 & phase-mixed & 6.81 & 1481 & 0.000 & False & 1375567 & MATCHED \\
6 & 3 & 1022 & stream & 6.80 & 1476 & 0.544 & True & 7306 & MATCHED \\
6 & 3 & 40713 & phase-mixed & 6.78 & 1362 & 0.000 & False & 608072 & MATCHED \\
6 & 3 & 86344 & stream & 6.70 & 1158 & 0.000 & True & 423599 & MATCHED \\
6 & 3 & 54423 & intact & 6.52 & 713 & 1.000 & False & 8695 & MATCHED \\
6 & 3 & 4317 & stream & 6.49 & 682 & 0.921 & False & 96349 & MATCHED \\
6 & 3 & 439497 & stream & 6.47 & 649 & 0.097 & False & 608129 & MATCHED \\
6 & 3 & 66116 & phase-mixed & 6.46 & 658 & 0.000 & False & 1969087 & MATCHED \\
6 & 3 & 86785 & phase-mixed & 6.45 & 631 & 0.000 & True & 510516 & MATCHED \\
6 & 3 & 450183 & phase-mixed & 6.42 & 577 & 0.000 & False & 1346695 & MATCHED \\
6 & 3 & 60914 & phase-mixed & 6.41 & 578 & 0.000 & False & 834323 & MATCHED \\
6 & 3 & 450996 & phase-mixed & 6.40 & 569 & 0.000 & False & 1346807 & MATCHED \\
6 & 3 & 820096 & phase-mixed & 6.35 & 505 & 0.000 & False & 661067 & MATCHED \\
6 & 3 & 119824 & stream & 6.21 & 362 & 0.000 & True & 510560 & MATCHED \\
6 & 3 & 401238 & phase-mixed & 6.19 & 344 & 0.000 & True & 485312 & MATCHED \\
6 & 3 & 818755 & phase-mixed & 6.15 & 318 & 0.000 & False & 786494 & MATCHED \\
6 & 3 & 22636 & phase-mixed & 6.14 & 315 & 0.000 & False & 709923 & MATCHED \\
6 & 3 & 83559 & phase-mixed & 6.09 & 279 & 0.000 & True & 485601 & MATCHED \\
6 & 3 & 820255 & phase-mixed & 6.07 & 267 & 0.000 & False & 7675517 & MATCHED \\
6 & 3 & 86573 & phase-mixed & 6.00 & 223 & 0.000 & True & 510832 & MATCHED \\
6 & 3 & 98912 & stream & 5.98 & 208 & 0.525 & False & 423105 & MATCHED \\
6 & 3 & 22641 & stream & 5.96 & 202 & 0.000 & False & 480459 & MATCHED \\
6 & 3 & 105073 & intact & 5.96 & 208 & 0.978 & False & 527547 & MATCHED \\
6 & 3 & 451182 & phase-mixed & 5.94 & 187 & 0.000 & False & 3975982 & MATCHED \\
6 & 3 & 25426 & stream & 5.84 & 161 & 0.067 & False & 439605 & MATCHED \\
6 & 3 & 450937 & phase-mixed & 5.83 & 147 & 0.000 & False & 1346868 & MATCHED \\
6 & 3 & 466589 & phase-mixed & 5.80 & 139 & 0.000 & False & 3971198 & MATCHED \\
6 & 3 & 461836 & phase-mixed & 5.74 & 116 & 0.000 & False & 4028766 & MATCHED \\
6 & 3 & 441640 & phase-mixed & 5.71 & 112 & 0.000 & False & 661038 & MATCHED \\
6 & 3 & 461788 & phase-mixed & 5.71 & 109 & 0.000 & False & 140 & MATCHED \\
6 & 3 & 60987 & phase-mixed & 5.70 & 109 & 0.000 & False & 1735450 & MATCHED \\
6 & 3 & 52829 & phase-mixed & 5.69 & 107 & 0.000 & True & 1719764 & MATCHED \\
\multicolumn{10}{c}{$\cdots$} \\
\hline
\end{tabular}